\documentclass[12pt]{article}
\usepackage{xcolor}
\usepackage{amsmath}
\usepackage[colorlinks=true,linkcolor=blue, citecolor=blue]{hyperref}%
\usepackage{amsfonts}
\usepackage{amssymb}
\usepackage{amsthm}
\usepackage{epsfig}
\usepackage{graphicx}
\usepackage[ruled,vlined]{algorithm2e}
\usepackage{hyperref}
\usepackage[all]{hypcap}
\usepackage{subfig}
\usepackage{graphicx}
\usepackage[countmax]{subfloat}
\usepackage{setspace}
\usepackage{subfig}
\usepackage{gensymb}
\usepackage{color}
\usepackage{bm}
\usepackage{authblk}
\usepackage{fullpage}
\usepackage{url}
\usepackage[authoryear]{natbib}
\usepackage{etoolbox}
\usepackage{amsthm}
\usepackage{multirow}
\usepackage[]{algorithm2e}
\usepackage{hyperref}

\theoremstyle{remark}

\makeatletter

\newif\ifabbreviation
\pretocmd{\thebibliography}{\abbreviationfalse}{}{}
\AtBeginDocument{\abbreviationtrue}

\newtheorem*{proposition1_1*}{Proposition 1}
\newtheorem*{proposition1_2*}{Proposition 2}
\newtheorem*{proposition1_3*}{Proposition 3}
\newtheorem*{proposition1_4*}{Proposition 4 (a discretized IDE-based representation)}
\newtheorem*{proposition1_5*}{Proposition 5}
\newtheorem*{proposition1_6*}{Proposition 6}
\newtheorem*{proposition1_7*}{Proposition 7}

\newtheoremstyle{exampstyle}
{4pt} 
{1pt} 
{} 
{} 
{\bfseries} 
{.} 
{.5em} 
{} 

\theoremstyle{exampstyle} 
\theoremstyle{exampstyle} 
\theoremstyle{exampstyle} 
\theoremstyle{exampstyle} 
\theoremstyle{exampstyle} 
\theoremstyle{exampstyle}

\usepackage{bm}
\usepackage{soul}
\usepackage{wrapfig}
\usepackage[]{algorithm2e}

\usepackage{titlesec}

\titlespacing\section{0pt}{12pt plus 4pt minus 2pt}{0pt plus 2pt minus 2pt}
\titlespacing\subsection{0pt}{12pt plus 4pt minus 2pt}{0pt plus 2pt minus 2pt}
\titlespacing\subsubsection{0pt}{12pt plus 4pt minus 2pt}{0pt plus 2pt minus 2pt}

\begin{document}


\title{Statistical Modeling for Spatio-Temporal Data from Stochastic Convection-Diffusion Processes}
\author[1]{Xiao Liu} 
\author[2]{Kyongmin Yeo} 
\author[2]{Siyuan Lu} 
\affil[1]{Department of Industrial Engineering\\ University of Arkansas}
\affil[2]{IBM T. J. Watson Research Center}


\maketitle

%
%


\singlespacing
\begin{abstract}

This paper proposes a physical-statistical modeling approach for spatio-temporal data arising from a class of stochastic convection-diffusion processes. Such processes are widely found in scientific and engineering applications where fundamental physics imposes critical constraints on how data can be modeled and how  models should be interpreted. 
The idea of spectrum decomposition is employed to approximate a physical spatio-temporal process by the linear combination of spatial basis functions and a multivariate random process of spectral coefficients. Unlike existing approaches assuming spatially- and temporally-invariant convection-diffusion, this paper considers a more general scenario with spatially-varying convection-diffusion and nonzero-mean source-sink. As a result, the temporal dynamics of spectral coefficients is coupled with each other, which can be interpreted as the non-linear energy redistribution across multiple scales from the perspective of physics. Because of the spatially-varying convection-diffusion, the space-time covariance is non-stationary in space.
The theoretical results are integrated into a hierarchical dynamical spatio-temporal model. 
The connection is established between the proposed model and the existing models based on Integro-Difference Equations.
Computational efficiency and scalability are also investigated to make the proposed approach practical. 
The advantages of the proposed methodology are demonstrated by numerical examples, a case study, and comprehensive comparison studies. Computer code is available on GitHub.  

\end{abstract}

\noindent\textbf{Key words:} {\em Convection-diffusion processes, spatio-temporal modeling, hierarchical dynamical spatio-temporal models, spectral method, radar-based precipitation nowcasting}

\clearpage
\singlespacing
\section{Introduction} \label{sec:one}
\noindent Digitization of the physical space has led to an explosive growth of spatio-temporal data from physical convection-diffusion processes. Examples span multi-disciplinary areas including environmental science, meteorology, remote sensing, geology, engineering, etc.
For such processes, fundamental physics  imposes critical constraints on how data can be modeled and how  models should be interpreted. This paper proposes a statistical modeling approach for spatio-temporal data arising from a generic class of convection-diffusion processes described by a Stochastic Partial Differential Equation (SPDE). By considering spatially-varying convection-diffusion, the proposed modeling approach makes critical extensions to the recent advances on the PDE-based statistical modeling of spatio-temporal data. 

\subsection{Motivation and Background}
Convection-diffusion equations have been widely used to describe how physical quantities (e.g., particles, air pollutant, energy, rain cells, heat, etc.) are transferred inside a physical system due to two processes: convection 
and diffusion. Interdisciplinary examples include, but are not limited to, urban air pollution processes, remote sensing of extreme weather systems such as hurricanes, decomposition of polymer in space and time, propagation of smoke from wildfires, regional epidemics of infectious diseases, etc.\citep{Stroud2010, Guinness2013, Sigrist2015, Liu2016, Lu2017, Liu2018, Guan2018}. 

For spatio-temporal data arising from physical convection-diffusion processes, deterministic physics-based numerical models are less effective for real-time operations due to the high computational cost and the ineffectiveness in utilizing real-time sensor data streams---a major source of information in the era of Big Data. 
For many complex physical processes, it is not only challenging to build the physics models from the first principles but also difficult to handle the high-degree uncertainty associated with model parameters and model inputs.  

Statistical spatio-temporal models, on the other hand, have been proven useful for short-term prediction and space-time interpolation using observations \citep{Wikle2001, Guinness2013, Liu2016, Kuusela2017, Reich2018, Guan2018}. For spatio-temporal data arising from physical processes, however, governing physics imposes critical constraints on how data can be modeled and interpreted. For example, the space-time covariance structures, to a great extent, are determined by the convection-diffusion equations from which the data are generated.
When the critical connection to fundamental physics is missing, the capabilities of pure data-driven approaches may not be fully appreciated by domain experts for monitoring, prediction and planning purposes.
In addition, for non-stationary and highly dynamic processes, the specification of statistically-adequate and physically-meaningful space-time covariance structures is known to be extremely challenging \citep{Nychka2002, Cressie2011, Simpson2012}.  Assumptions, such as isotropic, stationary and space-time separable, have been widely adopted which make the models mathematically tractable but too simple (or too abstract) to acknowledge the full complexity of the underlying physical processes. 

Hence, there is an urgent demand for intepretable models and computationally efficient algorithms tailored for large-scale spatio-temporal data arising from physical processes. 
When operational insights are required based on domain knowledge, our engagement with industry indicates that pure data-driven approaches could be limited even in data-rich scientific and engineering environments. As the volume of data increases, this issue appears to be more crucial than ever when statistical models are inevitably becoming more complex but seemingly less interpretable. Hence, data intensive physical analytics--an emerging field which resides at the intersection of physical modeling and statistics \citep{Hamann2016, Qian2019}--is being actively pursued in both academia and industry. 

\subsection{Literature Review}
The pioneering work of statistical modeling of spatio-temporal data can be found in \cite{Banerjee2004}, \cite{Schabenberger2005} and \cite{Cressie2011}. The mainstay approach models spatio-temporal processes by random fields with fully specified space-time covariance. This approach is referred to as the geostatistical paradigm and has gained tremendous popularity and success over the past decades. 
However, 
the specification of space-time covariance structures is known to be challenging for non-stationary and highly dynamic processes. 
Consider, for example, the modeling of urban air quality data collected by monitoring stations \citep{Liu2016}. The space-time correlation of the pollutant dispersion process depends on wind, temperature, solar radiation and traffic, and is too complex to be directly specified and validated. Similar examples can be found in the modeling of dynamic weather systems where the space-time covariance is non-stationary and depends on highly dynamic factors such as wind, small-scale localized convection, etc. \citep{Liu2018}. 
Hence, there have been prolonged research interests to provide flexible and effective ways to construct non-stationary covariance functions \citep{Cressie1999, Gneiting2002, Fuentes2005, Gneiting2006, Ghosh2010, Reich2011, Lenzi2019}. 
Moreover, the geostatistical modeling paradigm can be computationally expensive for large spatio-temporal data sets due to the prohibitive cubic operations $\mathcal{O}((TN)^3)$, where $T$ and $N$ are the number of samples in time and space \citep{Cressie2011, Banarjee2012, Simpson2012, Yan2017}. For example, a standard dual polarization meteorological Doppler weather radar image contains 230,400 pixels and new images are generated every 5 minutes. Due to the high computational cost, 
approximations are commonly used for large problems; for example, Gaussian Markov Random Fields representation \citep{Lindgren2011}, Nearest-Neighbor Gaussian Process \citep{ Datta2016, Banerjee2017}, kernel convolution \citep{Higdon1998}, low rank representation \citep{Cressie2002, Nychka2002, Banerjee2008}, approximation of likelihood functions \citep{Stein2004, Fuentes2007, Guinness2015}, Bayesian inference for latent Gaussian models based on the integrated nested Laplace
approximations \citep{Rue2009, R-Inla2019}, Lagrangian spatio-temporal covariance function \citep{Gneiting2006}, matrix-free state-space model \citep{Mondal2019}, Vecchia approximations of Gaussian processes \citep{Katzfuss2019},  as well as the multi-resolution approximation (\textit{M}-RA) of Gaussian processes observed at irregular spatial locations \citep{Katzfuss2017}. 

A powerful modeling framework, known as the Hierarchical Dynamical Spatio-Temporal Models (DSTM), has been proposed \citep{Wikle1999, Berliner2003, Cressie2011, Stroud2010, Katzfuss2019}. The power of the DSTM comes from the dynamical model specification by a series of conditional models that lead to a complex joint space-time covariance structures which can hardly be directly specified. \cite{Stroud2001} investigated a Gaussian state space framework where the mean function at each time is modeled by a locally weighted mixture of linear regressions. The use of Monte Carlo approaches makes the dynamical model more computable with very large data sets and for non-linear non-Gaussian models \citep{Carlin1992, Banerjee2004}. 

The SPDE-based modeling approach of large spatio-temporal data has also drawn much attention due to its connection to many physical convection-diffusion processes. 
\cite{Jone1997} developed statistical models for unequally spaced data arising from continuous stationary space-time processes described by a SPDE with white noise. Spectral density functions were obtained for the stationary space-time process and the inverse Fourier transform was employed to obtain the covariance functions. 
\cite{Brown2000} proposed a blur-generated non-separable space-time models and showed the explicit link of the proposed model to a convection-diffusion SPDE in the limiting case. 
\cite{Hooten2008} proposed a hierarchical Bayesian model for the spread of invasive species with spatially-varying diffusion coefficients as well as a logistic population growth term based on a common reaction-diffusion equation. 
\cite{Stroud2010} proposed a methodology for combining spatio-temporal satellite images with advection-diffusion models for interpolation and prediction of environmental processes.
\cite{Lindgren2011} investigated the explicit connection between a convection-diffusion SPDE and Gaussian Markov random field. 
\cite{Sigrist2015} proposed a systematic approach to obtain a Gaussian process by solving a SPDE with a convection-diffusion operator that does not vary in space and time. 
In the literature of computational mechanics, \cite{Qian2019} proposed a physics-informed method for learning low-dimensional models for large-scale dynamical systems govern by PDE. 
It is noted that a DSTM can also be motivated by special types of partial differential equations (e.g., diffusion, reaction-diffusion, or simplified convection-diffusion equations) under special conditions (e.g., spatially- and temporally-invariant convection-diffusion). A summary of the latest advances in the spatial modeling with SPDE can be found in \cite{Cressie2011} and \cite{Krainski2019}.

\subsection{Overview of the Paper}
This paper proposes a statistical modeling approach for spatio-temporal data arising from a generic class of convection-diffusion SPDE with a spatially-varying convection-diffusion operator and a nonzero-mean spatio-temporal source-sink term. 
We employ the important idea of spectrum decomposition to approximate the spatio-temporal process governed by a SPDE. Unlike the existing results based on spatially- and temporally-invariant convection-diffusion, the proposed model has two critical differences: (\textbf{i}) the temporal dynamics of spectrum coefficients is coupled when the convection-diffusion operator varies in space. Such a phenomenon is interpreted as the non-linear energy transfer or redistribution across multiple scales from the perspective of physics; (\textbf{ii}) while spatially- and temporally-invariant convection-diffusion leads to a stationary process in the limiting case, the proposed spatio-temporal model has a non-stationary space-time covariance structure due to the spatially-varying convection-diffusion. 
To our best knowledge, such a statistical modeling methodology is not yet available in the literature. 

In Section \ref{sec:general}, we describe the general modeling framework and present the key theoretical results. Section \ref{sec:special} focuses on real-valued processes under discrete sampling---a typical scenario in scientific and engineering applications. In Section \ref{sec:DSTM}, a systematic approach is presented to integrate the established theoretical results into the DSTM framework. Section \ref{sec:numeric} provides a numerical example, and a case study of radar-based precipitation nowcasting to demonstrate the advantages of the proposed method.

\section{The General Modeling Framework} \label{sec:general}
Consider a spatio-temporal convection-diffusion process governed by the following SPDE:
\begin{equation} \label{eq:SPDE1}
	\mathcal{A} \xi(t,\bm{s}) = Q(t,\bm{s}) + \varepsilon(t,\bm{s})
\end{equation}	
where $\xi(t,\bm{s})$ is a spatio-temporal process in space $\bm{s}$ and time $t$, $Q(t,\bm{s})$ is the source-sink term, $\varepsilon(t,\bm{s})$ is a spatio-temporal error process, and $\mathcal{A}$ is the convection-diffusion operator:
\begin{equation} \label{eq:A}
\mathcal{A}\xi(t,\bm{s}) = \frac{\partial}{\partial t} \xi(t,\bm{s}) + \bm{\vec{v}}_{t,\bm{s}}^T\triangledown \xi(t,\bm{s}) - \triangledown \cdot [\bm{D}_{t,\bm{s}} \triangledown  \xi(t,\bm{s})] + \zeta_{t, \bm{s}} \xi(t,\bm{s})
\end{equation}	
where $\bm{\vec{v}}_{t, \bm{s}}$, $\bm{D}_{t, \bm{s}}$, $\zeta_{t, \bm{s}}$, $\triangledown$ and $\triangledown\cdot$ represent the velocity, diffusivity, decay, gradient and divergence, respectively.

The SPDE (\ref{eq:SPDE1}) describes the transport of particles, energy, or other physical quantities by convection and diffusion processes.
The second term on the right hand side of (\ref{eq:A}) models the convection of a physical quantify by an ambient flow, while the third and the last terms represent diffusion and decay, respectively. Depending on the applications, (\ref{eq:SPDE1}) is referred to as the \textit{convection-diffusion}, \textit{advection-diffusion}, \textit{drift-diffusion}, or \textit{scalar transport} equations.
When applied to model real data, the SPDE (\ref{eq:SPDE1}) is not exactly satisfied due to the uncertainties in the model parameters. For example, the measurement errors or noises introduce aleatoric uncertainties in the velocity $\bm{\vec{v}}_{t,\bm{s}}$ and diffusivity $\bm{D}_{t,\bm{s}}$ (see Proposition 1), while our incomplete knowledge on the forcing term, $Q(t,\bm{s})$, often leads to epistemic uncertainties (see the case study in Section 5). 
Hence, by incorporating the error process $\varepsilon(t,\bm{s})$ into the SPDE (\ref{eq:SPDE1}), one can exploit the capability of statistical approaches to quantify the sources of uncertainty for complex physical processes. 

The spectral decomposition of $\xi(t,\bm{s})$ is given by the linear combination of spatial basis functions and random spectral coefficients \citep{Wikle1999, Cressie2011, Sigrist2015, Mak2018, Qian2019}:

\begin{equation} \label{eq:decomposition}
\xi(t,\bm{s}) \approx \tilde{\xi}(t,\bm{s}) = \sum_{j=1}^{K}\alpha_j(t)f_j(\bm{s}) \equiv  \bm{f}^T(\bm{s})\bm{\alpha}(t).
\end{equation}

Here, $\bm{\alpha}(t) = (\alpha_1(t),...,\alpha_K(t) )^T$ is a multivariate temporal process of spectral coefficients, $f_j(\bm{s})=\exp(i\bm{k}_j^T\bm{s})$ is the deterministic Fourier basis function, $\bm{k}_j$ is the spatial wavenumber, and $\bm{f}(\bm{s}) = (f_1(\bm{s}),...,f_K(\bm{s}) )^T$. The choice of Fourier spatial basis function is appropriate if $\xi(t,\bm{s})$ is periodic.   
The spectral decomposition (\ref{eq:decomposition}) provides some key advantages in modeling the spatio-temporal data arising from the physical process (\ref{eq:SPDE1}):

\begin{itemize}
	\item Once the spatial basis functions have been fixed, the modeling of a spatio-temporal process $\xi(t,\bm{s})$ can be converted to the modeling of a multivariate stochastic process $\bm{\alpha}(t)$. The latter problem is often more tractable.
	\item If it is possible to find a multivariate stochastic process $\bm{\alpha}(t)$ such that the approximated process generated by (\ref{eq:decomposition}) also satisfies the SPDE (\ref{eq:SPDE1}), the fundamental physics in (\ref{eq:SPDE1}) can be naturally built into the statistical model \citep{Wikle1999, Sigrist2015, Qian2019}. Alternatively, one may choose to model $\bm{\alpha}(t)$ by pure data-driven approaches, if incorporating physics into statistical models is not a major modeling concern \citep{Yeo2018}.  
	\item Once the temporal dynamics of $\bm{\alpha}(t)$ has been established, the space-time covariance structure of $\tilde{\xi}(t,\bm{s})$ can be derived from the spectral decomposition (\ref{eq:decomposition}), where $\tilde{\xi}(t,\bm{s})$ is a linear combination of $\bm{\alpha}(t)$ and known basis functions $\bm{f}(\bm{s})$. 
	\item Dimension reduction is possible by retaining only the low-frequency components in (\ref{eq:decomposition}), which helps to achieve a tremendous computational advantage without significantly sacrificing the modeling accuracy. This is often a critical step that ensures the proposed model to be practical for certain applications as shown in the case study.   
	\item The model based on (\ref{eq:decomposition}) can be integrated into the framework of DSTM \citep{Wikle1998, Berliner2003, Cressie2011}, which enables computationally efficient algorithms for statistical inference, monitoring and prediction. 
\end{itemize}

We firstly show, in Proposition 1, that a multivariate stochastic process $\bm{\alpha}(t)$ does exist such that the approximated process generated by (\ref{eq:decomposition}) satisfies the SPDE (\ref{eq:SPDE1}). 
The following assumptions are made: (\textit{A1}) the convection-diffusion operator $\mathcal{A}$ in (\ref{eq:A}) is spatially-varying but temporally-invariant  (i.e., $\bm{\vec{v}}_{t,\bm{s}}=\bm{\vec{v}}_{\bm{s}}$, $\bm{D}_{t,\bm{s}}=\bm{D}_{\bm{s}}$, $\zeta_{t, \bm{s}}=\zeta_{\bm{s}}$); (\textit{A2}) the source-sink process $Q(t,\bm{s})$  admits a spectral representation $Q(t,\bm{s})  = \bm{f}^T(\bm{s})\bm{\beta}(t)$ with $\bm{\beta}(t) = (\beta^{(1)}(t),...,\beta^{(K)}(t) )^T$ dynamically evolving over time; (\textit{A3}) the error process $\varepsilon(t,\bm{s})=\bm{f}^T(\bm{s})\tilde{\bm{\varepsilon}}(t)$ is a white-in-time stationary spatial process where $\tilde{\bm{\varepsilon}}(t)$ is a $K$-dimensional random vector; (\textit{A4}) the initial condition is given by $\tilde{\xi}(0,\bm{s})  = \bm{f}^T(\bm{s})\bm{\alpha}(0)$ where $\bm{\alpha}(0)\sim N(0,\mathrm{diag}\{\tilde{h}_0(\bm{k}_j) \})$ and $\tilde{h}_0(\bm{k}_j)$ is the spectral density.

\begin{proposition1_1*}
\textit{For spatially-varying convection-diffusion (i.e., \textit{A1}), if the initial condition $\tilde{\xi}(0,\bm{s})$, source-sink process $Q(t,\bm{s})$ and error process $\varepsilon(t,\bm{s})$ are in a space $\mathbb{S}$ spanned by a finite number of Fourier functions $\bm{f}^T(\bm{s})$ (i.e., \textit{A2$\sim$A4}), then, there exists a multivariate temporal process $\bm{\alpha}(t)$ such that $\tilde{\xi}(t,\bm{s})$ remains in $\mathbb{S}$ and satisfies the SPDE (\ref{eq:SPDE1}). The process $\bm{\alpha}(t)$ is characterized by a stochastic Ordinary Differential Equation (ODE) as follows:
	\begin{equation} \label{eq:gamma_transition_2}
	\dot{\bm{\alpha}}(t) =  \bm{G}\bm{\alpha}(t) + \bm{\beta}(t) + \bm{\tilde{\varepsilon}}(t).
	\end{equation}
	\indent Here, the $(i,j)$th entry of the non-diagonal matrix $\bm{G}$ is given by
		\begin{equation}
	g_{i,j}= C_j^{-1} \int_{\mathbb{S}}\left\{-\bm{k}^T_j \bm{D}_{\bm{s}}\bm{k}_j - \zeta_{\bm{s}} - \imath  (\bm{\vec{v}}_{\bm{s}}^T \bm{k}_j - [\triangledown\cdot \bm{D}_{\bm{s}}]^T\bm{k}_j)    \right\} f_j{(\bm{s})}f_i^*{(\bm{s})}d\bm{s}
	\end{equation}
		where $\imath$ is the imaginary unit, $C_j=\int f_j(\bm{s})f_j^*(\bm{s})d\bm{s}$, $\cdot ^ *$ represents the complex conjugation, and the covariance of the error process $\bm{\tilde{\varepsilon}}(t)$ is given by $\mathrm{cov}(\tilde{\bm{\varepsilon}}(t),\tilde{\bm{\varepsilon}}(t'))=\delta_{t,t'}\mathrm{diag}\{\tilde{h}(\bm{k}_j)\}$ with $\delta_{t,t'}$ and $\tilde{h}(\cdot)$ respectively being the Kronecker delta and spectral density function.}
\end{proposition1_1*}


\textit{\textbf{Remark.}} The proof of Proposition 1, using the Galerkin method, is provided in Appendix A. The temporal evolution of $\bm{\alpha}(t)$ described by the ODE (\ref{eq:gamma_transition_2}) has interesting interpretations: the first term on the right hand side, $\bm{G}\bm{\alpha}(t)$, captures the transition of the spectral coefficients $\bm{\alpha}(t)$ due to the convection-diffusion operator $\mathcal{A}$; the second term, $\bm{\beta}(t)$, captures the additional change to $\bm{\alpha}(t)$ due to the source-sink $Q(t,\bm{s})$; and the third term $\tilde{\bm{\varepsilon}}(t)$ captures the uncertainty associated with the evolution of $\bm{\alpha}(t)$ due to the error process $\varepsilon(t,\bm{s})$ in (\ref{eq:SPDE1}).

Note that the process $\bm{\alpha}(t)$ defined in (\ref{eq:gamma_transition_2}) differs from the existing result in the literature. Since $\bm{G}$ is non-diagonal given a spatially-varying convection-diffusion operator $\mathcal{A}$, the ODE (\ref{eq:gamma_transition_2}) immediately implies that the temporal evolution of the components in $\bm{\alpha}(t)$ are coupled: \textit{at any time $t$, the transition of each component in $\bm{\alpha}(t)$ depends not only on itself, but also on all other components in $\bm{\alpha}(t)$}. If the Fourier coefficients are viewed as the energy distributed to different frequencies, the coupling of Fourier coefficients can be naturally interpreted as the \textit{energy transfer across multiple scales}, i.e., the spatially-varying convection-diffusion redistributes the energy across different frequencies. 
Under a special case when the convection-diffusion operator does not vary in space and time, and the source-sink is negligible (i.e., $Q(t,\bm{s})=0$, $\bm{\vec{v}}_{t,\bm{s}}=\bm{\vec{v}}$ and $\bm{D}_{t,\bm{s}}=\bm{D}$), the ODE (\ref{eq:gamma_transition_2}) degenerates to the result in \cite{Sigrist2015} where the temporal evolution of the components in $\bm{\alpha}(t)$ are completely de-coupled (i.e., $\bm{G}$ becomes diagonal), leading to a stationary solution of (\ref{eq:SPDE1}) in the limiting case; also see \cite{Brown2000}. When the velocity field $\bm{\vec{v}}_{\bm{s}}$ and diffusivity $\bm{D}_{\bm{s}}$ vary in space, however, the process considered in our model becomes non-stationary in space.


Once the temporal dynamics of $\bm{\alpha}(t)$ has been established, the space-time covariance of $\tilde{\xi}(t,\bm{s})$ is derived from the spectral decomposition (\ref{eq:decomposition}), where $\tilde{\xi}(t,\bm{s})$ is a linear combination of $\bm{\alpha}(t)$ and fixed basis functions $\bm{f}(\bm{s})$. It follows from (\ref{eq:decomposition}) that
\begin{equation} 
\mathrm{cov}(\tilde{\xi}(t+\Delta,\bm{s}), \tilde{\xi}(t,\bm{s}')) = \bm{f}^T(\bm{s})\mathrm{cov}(\bm{\alpha}(t+\Delta),\bm{\alpha}(t))\bm{f}(\bm{s}').
\end{equation}
Hence, it is sufficient to obtain the covariance structure of the multivariate temporal process $\bm{\alpha}(t)$ given by the following proposition. 

\vspace{0.2in}
\begin{proposition1_2*} 
\textit{For the multivariate temporal process $\bm{\alpha}(t)$ defined in (\ref{eq:gamma_transition_2}), we have 
	\begin{equation} \label{eq:alpha_cov}
	\mathrm{cov}(\bm{\alpha}(t+\Delta), \bm{\alpha}(t)) = 	\exp(\bm{G}\Delta)\left( \exp(\bm{G}t)\bm{H}_0\exp^*(\bm{G}^Tt) + \int_{0}^{t}\exp(\bm{G}(t-\tau))\bm{H}\exp^*(\bm{G}^T(t-\tau))d\tau \right) 	
\end{equation}
where $\bm{H}=\mathrm{diag}(\tilde{h}(\bm{k}_j))$ and $\bm{H}_0=\mathrm{diag}(\tilde{h}_0(\bm{k}_j))$ are respectively the spectral density of $\bm{\tilde{\varepsilon}}(t)$ and $\bm{\alpha}(0)$. When $t\rightarrow\infty$, the effect of the initial condition $\bm{\alpha}(0)$ vanishes and we have 
\begin{equation} 
\lim_{t\rightarrow \infty} \mathrm{cov}(\bm{\alpha}(t+\Delta), \bm{\alpha}(t)) = 	\exp(\bm{G}\Delta) \int_{0}^{t}\exp(\bm{G}(t-\tau))\bm{H}\exp^*(\bm{G}^T(t-\tau))d\tau.
\end{equation}}
\end{proposition1_2*}


\textit{\textbf{Remark.}} The derivation of (\ref{eq:alpha_cov}) is provided in Appendix B. To generate some insights on Proposition 2, we first consider the discrete-time representation of the ODE (\ref{eq:gamma_transition_2}), $\bm{\alpha}(t+\Delta)  = \exp(\bm{G}\Delta)\bm{\alpha}(t) + \bm{q}(\Delta)$, where
\begin{equation}
\bm{q}(\Delta) \sim N \left(  \int_{t}^{t+\Delta}\bm{\beta}(t)dt, \int_{0}^{\Delta}\exp(\bm{G}(\Delta-\tau))\bm{H}\exp^*(\bm{G}^T(\Delta-\tau))d\tau\right)
\end{equation}
represents the amount of shift applied to $\bm{\alpha}(t)$ over a time interval $\Delta$ due to both the source-sink and error processes in (\ref{eq:SPDE1}). Then, under a special case with constant convection, diffusion and decay (i.e., $\bm{\vec{v}}_{\bm{s}}=\bm{\vec{v}}$, $\bm{D}_{\bm{s}}=\mathrm{diag}(d)$ and $\zeta_{\bm{s}}=\zeta$), $\bm{G}$ becomes a diagonal matrix with its $(i,i)$th entry being given by $g_{i,i}=-d\bm{k}_i^T\bm{k}_i-\zeta-\imath \bm{\vec{v}}^{T}_{\bm{s}}\bm{k}_i$, and the $(i,i)$th entry of the covariance matrix of  $\bm{q}(\Delta)$ becomes
\begin{equation} \label{eq:s_q}
\begin{split}
\tilde{h}(\bm{k}_i) & \int_{0}^{\Delta}\exp(-(d\bm{k}_i^T\bm{k}_i + \zeta+\imath \bm{\vec{v}}^{T}_{\bm{s}}\bm{k}_i)(\Delta-\tau))\exp^*(-(d\bm{k}_i^T\bm{k}_i + \zeta+\imath \bm{\vec{v}}^{T}_{\bm{s}}\bm{k}_i)(\Delta-\tau))d\tau \\
& =\frac{\tilde{h}(\bm{k}_i)}{2d\bm{k}_i^T\bm{k}_i + 2\zeta}\left[1-\exp(-(2d\bm{k}_i^T\bm{k}_i + 2\zeta)\Delta)  \right].
\end{split}
\end{equation}

Equation (\ref{eq:s_q}) indicates that: (i) the uncertainty associated with $\bm{q}(\Delta)$ does not depend on convection; (ii) for small $\Delta$, $d$ and $\zeta$, (\ref{eq:s_q}) can be approximated by $\tilde{h}(\bm{k}_i)\Delta$ which indicates that the effects of diffusion and decay are small on $\bm{q}(\Delta)$ within a small time interval $\Delta$; (iii) when $\Delta\rightarrow\infty$, (\ref{eq:s_q}) becomes $\tilde{h}(\bm{k}_i)(2d\bm{k}_i^T\bm{k}_i + 2\zeta)^{-1}$ which indicates that the diffusion and decay force the spectrum of the noise to decay. 

\section{Real-Valued Processes on Space-Time Grids} \label{sec:special}
In Section \ref{sec:special}, we restrict our attention to an important scenario in scientific and engineering applications: real-valued spatio-temporal processes on space-time grids. 
For example, the temperature field in a data center server room is sampled by thermal sensors at discrete spatial locations and time intervals, radar or remote sensing images are taken at fixed time intervals and over fixed-size image pixels, etc.

We consider a real-valued process $\xi(t,\bm{s})$ defined on a $N_1 \times N_2$ regular spatial grid and at time points $t_1,t_2,...,t_T$. In particular, the rectangular spatial mesh system is defined by a tensor product of two one-dimensional collocation sets,
\begin{equation} \label{eq:tensor}
\mathbb{S} = \bm{s}_1 \otimes \bm{s}_2, 
\end{equation}
where $\bm{s}_1$ and $\bm{s}_2$ are $\bm{s}_1 = (s_1^{(i)}; s_1^{(i)} = \frac{i}{N_1}, i = 0,\cdots,N_1-1)$ and $\bm{s}_2 = (s_2^{(i)}; s_2^{(i)} = \frac{i}{N_2}, i = 0,\cdots,N_2-1)$.
Without loss of generality, $N_1$ and $N_2$ are assumed to be even numbers. 

\subsection{2D Discrete Fourier Transform} \label{sec:2DDFT}
The proposed approach is deeply rooted in the 2D discrete Fourier transform (DFT). 
For a 2D process $x(n_1, n_2)$ at grid points $(n_1, n_2)$, $n_1=0,2,...,N_1-1$ and $n_2=0,2,...,N_2-1$, the DFT of $x(n_1, n_2)$ is defined as
\begin{equation}  \label{eq:DFT}
\begin{split}
X(k_1,k_2) & = \frac{1}{N_1N_2}\sum_{n_1=0}^{N_1-1}\sum_{n_2=0}^{N_2-1} x(n_1, n_2) e^{-\imath (\frac{2\pi}{N_1}n_1k_1+\frac{2\pi}{N_2}n_2k_2)} \\
& = \frac{1}{N_1N_2}\sum_{n_1}\sum_{n_2}x(n_1, n_2)\cos(\frac{2\pi}{N_1}n_1k_1+\frac{2\pi}{N_2}n_2k_2))\\& \quad - \imath \frac{1}{N_1N_2}\sum_{n_1}\sum_{n_2}x(n_1, n_2)\sin(\frac{2\pi}{N_1}n_1k_1+\frac{2\pi}{N_2}n_2k_2)  \equiv \alpha^{(R)}_{k_1,k_2} - \imath  \alpha^{(I)}_{k_1,k_2}
\end{split}
\end{equation}
for $k_1 = -N_1/2+1, -N_1/2+2, ..., N_1/2$ and  $k_2=-N_2/2+1, -N_2/2+2, ..., N_2/2$. The inverse transform is also obtained as:
\begin{equation} \label{eq:inv_DFT}
\begin{split}
x&(n_1, n_2)  = \sum_{k_1=-N_1/2+1}^{N_1/2}\sum_{k_2=-N_2/2+1}^{N_2/2} X(k_1,k_2) e^{\imath (\frac{2\pi}{N_1}n_1k_1+\frac{2\pi}{N_2}n_2k_2)} \\
& =  \sum_{k_1}\sum_{k_2} (\alpha^{(R)}_{k_1,k_2} - \imath  \alpha^{(I)}_{k_1,k_2}) \left\{\cos(\frac{2\pi}{N_1}n_1k_1+\frac{2\pi}{N_2}n_2k_2)+\imath \sin(\frac{2\pi}{N_1}n_1k_1+\frac{2\pi}{N_2}n_2k_2) \right\} \\
& = \sum_{k_1}\sum_{k_2}\alpha^{(R)}_{k_1,k_2}\cos(\frac{2\pi}{N_1}n_1k_1+\frac{2\pi}{N_2}n_2k_2) + \sum_{k_1}\sum_{k_2}\alpha^{(I)}_{k_1,k_2}\sin(\frac{2\pi}{N_1}n_1k_1+\frac{2\pi}{N_2}n_2k_2).
\end{split}
\end{equation}
For real-valued $x(n_1, n_2)$, it is well known that (i.e., rotational symmetry)
\begin{equation} \label{eq:duality}
|X(k_1,k_2)|=|X(-k_1,-k_2)|,\quad \arg X(k_1,k_2) = -\arg X(-k_1,-k_2).
\end{equation}

\begin{figure}[h!]  
	\begin{center}
		\includegraphics[width=0.9\textwidth]{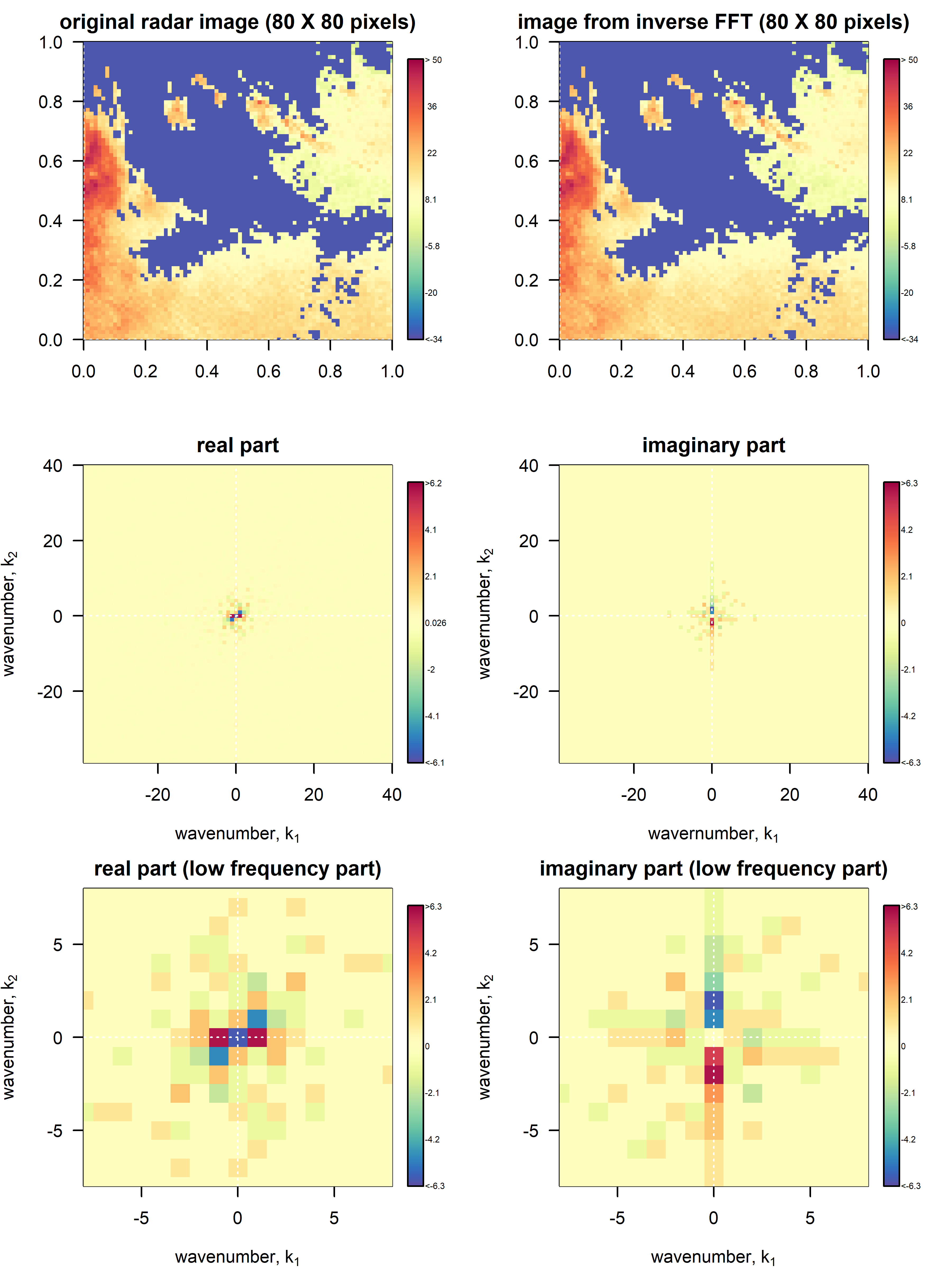}
		\centering
		\caption{Illustrations of 2D DFT for real-valued processes (top row: the original radar image and the recovered image from the inverse DFT; mid row: the real and imaginary parts of the FFT, $X(k_1, k_2)$; bottom row: the low-frequency region of the real and imaginary parts of $X(k_1, k_2)$, i.e., a zoomed-in view).}
		\label{fig:fig2}
	\end{center}
\end{figure}
\clearpage

As an illustration, Figure \ref{fig:fig2} shows a $80 \times 80$ pixel weather radar image and its 2D DFT. 
The first row of this figure shows the original image (left) and the recovered image (right) from the inverse DFT; the second row shows the real and imaginary parts of the FFT, $X(k_1, k_2)$, for $k_1,k_2=-39,-38,...,40$; and the bottom row presents a zoomed-in view of the real and imaginary parts of $X(k_1, k_2)$ by only keeping the low-frequency components. For such a real-valued process, the rotational symmetry of the real and imaginary parts of $X(k_1, k_2)$ is clearly seen, i.e., $|X(k_1,k_2)|=|X(-k_1,-k_2)|$ and $\arg X(k_1,k_2) = -\arg X(-k_1,-k_2)$. One key observation from Figure \ref{fig:fig2} is that the Fourier coefficients corresponding to the high-frequency terms (approximately $k>15$) are practically zero, which strongly suggests the possibility of dimension reduction by keeping only the low-frequency components. If it is assumed that the image in Figure \ref{fig:fig2} is band-limited within $[-15,15]$, then, the spatial sampling rate in this example (i.e., $1/80$) far exceeds the Nyquist rate, which explains why the image can be well reconstructed by inverse FFT.

\subsection{The Statistical Model for Real-Valued Processes}
The 2D DFT motivates us to approximate a real-valued spatio-temporal process $\xi(t,\bm{s})$ by allowing $\alpha^{(R)}_{k_1,k_2}$ and $\alpha^{(I)}_{k_1,k_2}$ to vary over time. Hence, we have
\begin{equation}  \label{eq:model_DFT}
\xi(t,\bm{s}) \approx \sum_{k_1=-N_1/2+1}^{N_1/2}\sum_{k_2=-N_2/2+1}^{N_2/2} \alpha_{k_1,k_2}^{(R)}(t)f_{k_1,k_2}^{(R)}(\bm{s})+ \alpha_{k_1,k_2}^{(I)}(t)f_{k_1,k_2}^{(I)}(\bm{s}) 
\end{equation}
where $f_{k_1,k_2}^{(R)}(\bm{s})=\cos(2\pi s_1k_1+2\pi s_2k_2)$, $f_{k_1,k_2}^{(I)}(\bm{s})=\sin(2\pi s_1k_1+2\pi s_2k_2)$. Recall that, $\xi(t,\bm{s})$ is a real-valued process on the rectangular mesh system (\ref{eq:tensor}) at time points $t_1,t_2,...,t_T$. 

Because of the rotational symmetry (\ref{eq:duality}), we obtain the following constraints:
\begin{equation} \label{eq:alpha_con1}
\alpha_{k_1,k_2}^{(R)}(t) = \alpha_{-k_1+iN_1,-k_2+jN_2}^{(R)}(t), \alpha_{k_1,k_2}^{(I)}(t) = -\alpha_{-k_1+iN_1,-k_2+jN_2}^{(I)}(t), \quad i,j \in \mathbb{N}
\end{equation}
and
\begin{equation} \label{eq:alpha_con2}
\alpha_{0,0}^{(I)}(t) = \alpha_{0,\frac{N_2}{2}}^{(I)}(t) = \alpha_{\frac{N_1}{2},0}^{(I)}(t) = \alpha_{\frac{N_1}{2},\frac{N_2}{2}}^{(I)}(t) = 0.
\end{equation}

The constraints  (\ref{eq:alpha_con1}) and (\ref{eq:alpha_con2}) enable us to re-write (\ref{eq:model_DFT}) as
\begin{equation} \label{eq:model_DFT_2}
\xi(t,\bm{s}) = \sum_{\bm{k}\in\Omega_1}\alpha_{\bm{k}}^{(R)}(t)f_{\bm{k}}^{(R)}(\bm{s})
+ 2\sum_{\bm{k}\in\Omega_2}\left(\alpha_{\bm{k}}^{(R)}(t)f_{\bm{k}}^{(R)}(\bm{s})+\alpha_{\bm{k}}^{(I)}(t)f_{\bm{k}}^{(I)}(\bm{s})\right)
\end{equation}
where $\bm{k}=(k_1,k_2)$, and the two sets of spatial wavenumbers, $\Omega_1$ and $\Omega_2$, are defined as:
\begin{equation}
\begin{split}
& \Omega_1 = \{(0,0), (0,\frac{N_2}{2}), (\frac{N_1}{2},0), (\frac{N_1}{2},\frac{N_2}{2})\} \\
& \Omega_2 = \{(k_1,k_2); k_1=0,1,...,\frac{N_1}{2},k_2=0,1,...,\frac{N_2}{2}\} \\
& \quad \quad \quad \cup \{(k_1,k_2); k_1=-1,...,-\frac{N_1}{2}+1,k_2=-1,...,-\frac{N_2}{2}+1\} \setminus \Omega_1.
\end{split}
\end{equation}

It is easy to see that the dimension of the Fourier coefficients is $N_1 \times N_2$, i.e., the total number of $\alpha^{(R)}$ and $\alpha^{(I)}$ to be determined in (\ref{eq:model_DFT_2}). In spectral analysis, it is a common practice to further drop the terms corresponding to the highest frequencies, $\frac{N_1}{2}$ and $\frac{N_2}{2}$, in order to achieve the rotational symmetry between all pairs in the wavenumber space. Hence, an alternative approach to (\ref{eq:model_DFT_2}) is:
\begin{equation} \label{eq:model_DFT_3}
\xi(t,\bm{s}) = \left[ \alpha_{\bm{k}}^{(R)}(t)f_{\bm{k}}^{(R)}(\bm{s})\right]_{\bm{k}=(0,0)}
+ 2\sum_{\bm{k}\in\Omega_3}\left(\alpha_{\bm{k}}^{(R)}(t)f_{\bm{k}}^{(R)}(\bm{s})+\alpha_{\bm{k}}^{(I)}(t)f_{\bm{k}}^{(I)}(\bm{s})\right)
\end{equation}
where 
\begin{equation}
\begin{split}
\Omega_3 = & \{(k_1,k_2); k_1=0,1,...,\frac{N_1}{2}-1,k_2=0,1,...,\frac{N_2}{2}-1\} \\
& \quad \quad \quad \cup \{(k_1,k_2); k_1=-1,...,-\frac{N_1}{2}+1,k_2=-1,...,-\frac{N_2}{2}+1\} \setminus \ \{(0,0)\}.
\end{split}
\end{equation}
and the dimension of (\ref{eq:model_DFT_3}) is reduced to $N_1\times N_2 - N_1 - N_2 + 1$. For example, if $N_1 = N_2 = 80$ as shown in Figure \ref{fig:fig2}, the dimensions of (\ref{eq:model_DFT_2}) and (\ref{eq:model_DFT_3}) are respectively 6,400 and 6,241. It is also shown in Figure \ref{fig:fig2} that the Fourier coefficients corresponding to the high-frequency components are close to zero (i.e., the energy is concentrated in the low-frequency region). Hence, the approximation (\ref{eq:model_DFT_3}) is well justified. 

\begin{figure}[h!]
	\begin{center}
		\includegraphics[width=1\textwidth]{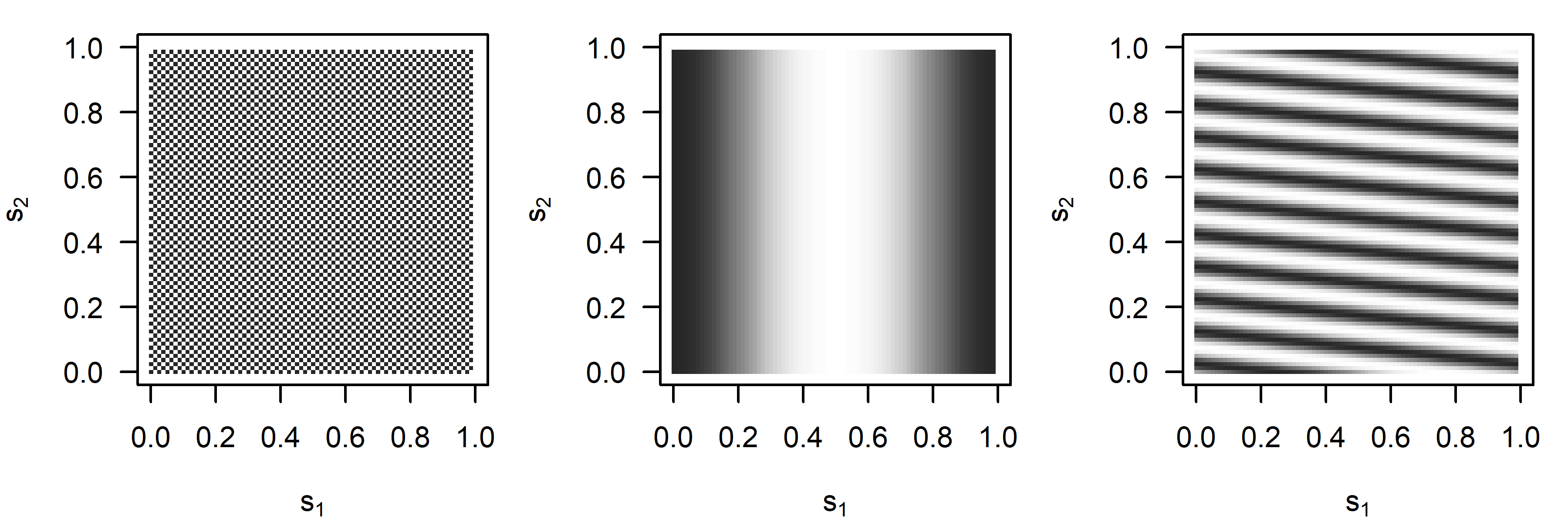}
		\centering
		\caption{Fourier basis functions at selected frequencies:  $f_{40,40}^{(R)}(\bm{s})$,$f_{1,0}^{(R)}(\bm{s})$,and $f_{1,10}^{(I)}(\bm{s})$.}
		\label{fig:fig3} 
	\end{center}
\end{figure}
For the same radar image shown in Figure \ref{fig:fig2}, Figure \ref{fig:fig3} illustrates the Fourier basis functions at selected frequencies. The left panel shows $f_{40,40}^{(R)}(\bm{s})$ with $k_1=k_2=40$ (the last entry in $\Omega_1$), the middle panel shows $f_{1,0}^{(R)}(\bm{s})$ with $k_1=1$ and $k_2=0$ (the 40$^{th}$ entry in $\Omega_2$), and the right panel shows $f_{1,10}^{(I)}(\bm{s})$ with $k_1=1$ and $k_2=10$ (the 50$^{th}$ entry in $\Omega_2$).

Similar to Proposition 1, Proposition 3 establishes the temporal dynamics of $\alpha_{\bm{k}}^{(R)}(t)$ and $\alpha_{\bm{k}}^{(I)}(t)$ such that the approximated process (\ref{eq:model_DFT_2}) satisfies the SPDE (\ref{eq:SPDE1}). In the case that the process is approximated by (\ref{eq:model_DFT_3}), the extension of Proposition 3 is trivial and thus omitted. 

\begin{proposition1_3*}
	\textit{Let $\bm{\alpha}(t) = (  \{\alpha_{\bm{k}}^{(R)}(t);\bm{k}\in\Omega_1\}, \{\alpha_{\bm{k}}^{(R)}(t);\bm{k}\in\Omega_2\}, \{\alpha_{\bm{k}}^{(I)}(t);\bm{k}\in\Omega_2\} )^T$ be a collection of spectral coefficients; let $\tilde{\xi}(0,\bm{s})$ be the initial condition where $\bm{\alpha}(0)\sim N(0, \bm{H}_0 )$ with $\bm{H}_0 = \mathrm{diag}(\tilde{h}_0(\bm{k}))$ being a spectral density; let}
	\begin{equation}
	Q(t,\bm{s}) = \sum_{\bm{k}\in\Omega_1}\beta_{\bm{k}}^{(R)}(t)f_{\bm{k}}^{(R)}(\bm{s})
	+ 2\sum_{\bm{k}\in\Omega_2}\left\{\beta_{\bm{k}}^{(R)}(t)f_{\bm{k}}^{(R)}(\bm{s})+\beta_{\bm{k}}^{(I)}(t)f_{\bm{k}}^{(I)}(\bm{s})\right\}
	\end{equation}
	\textit{be the spectral representation of the deterministic source-sink process; let}
	\begin{equation}
	\varepsilon(t,\bm{s}) = \sum_{\bm{k}\in\Omega_1}\tilde{\bm{\varepsilon}}_{\bm{k}}^{(R)}(t)f_{\bm{k}}^{(R)}(\bm{s})
	+ 2\sum_{\bm{k}\in\Omega_2}\left\{\tilde{\bm{\varepsilon}}_{\bm{k}}^{(R)}(t)f_{\bm{k}}^{(R)}(\bm{s})+\tilde{\bm{\varepsilon}}_{\bm{k}}^{(I)}(t)f_{\bm{k}}^{(I)}(\bm{s})\right\}
	\end{equation}
	\textit{be the spectral representation of the white-in-time stationary error process, and let $\tilde{\bm{\varepsilon}}(t)=(  \{\tilde{\bm{\varepsilon}}^{(R)}(t);\bm{k}\in\Omega_1\}, \{\tilde{\bm{\varepsilon}}^{(R)}(t);\bm{k}\in\Omega_2\}, \{\tilde{\bm{\varepsilon}}^{(I)}(t);\bm{k}\in\Omega_2\} )^T$ be a random noise vector with spectral density $\bm{H} =\mathrm{diag}(\tilde{h}(\bm{k}))$. Then, the temporal dynamics of $\alpha_{\bm{k}}^{(R)}(t)$ and $\alpha_{\bm{k}}^{(I)}(t)$ is given as follows such that the approximated process (\ref{eq:model_DFT_2}) satisfies the real-valued SPDE (\ref{eq:SPDE1}):}
	
	\textit{For $\bm{k}' \in \Omega_1$,} 
	\begin{equation} \label{eq:p3_1}
	\begin{split}
	\dot{\alpha}_{\bm{k}'}^{(R)}(t) = &
	C_{\bm{k}}^{-1}\sum_{\bm{k}\in\Omega_1} \alpha_{\bm{k}}^{(R)}(t)  \Psi_{1,5,9}(\bm{k},\bm{k}')+  2C_{\bm{k}}^{-1}\sum_{\bm{k}\in\Omega_2} \alpha_{\bm{k}}^{(R)}(t)  \Psi_{1,5,9}(\bm{k},\bm{k}') \\ & +  2C_{\bm{k}}^{-1}\sum_{\bm{k}\in\Omega_2} \alpha_{\bm{k}}^{(I)}(t)  \Psi_{2,6,10}(\bm{k},\bm{k}') + \beta_{\bm{k}'}^{(R)}(t) + \tilde{\bm{\varepsilon}}_{\bm{k}'}^{(R)}(t)
	\end{split}	
	\end{equation}
	
	\textit{For $\bm{k}' \in \Omega_2$,} 
	\begin{equation} \label{eq:p3_2}
	\begin{split}
	\dot{\alpha}_{\bm{k}'}^{(R)}(t) = &
	(2C_{\bm{k}})^{-1}\sum_{\bm{k}\in\Omega_1} \alpha_{\bm{k}}^{(R)}(t)  \Psi_{1,5,9}(\bm{k},\bm{k}')+  C_{\bm{k}}^{-1}\sum_{\bm{k}\in\Omega_2} \alpha_{\bm{k}}^{(R)}(t)  \Psi_{1,5,9}(\bm{k},\bm{k}') \\ & +  C_{\bm{k}}^{-1}\sum_{\bm{k}\in\Omega_2} \alpha_{\bm{k}}^{(I)}(t)  \Psi_{2,6,10}(\bm{k},\bm{k}') + \beta_{\bm{k}'}^{(R)}(t) + \tilde{\bm{\varepsilon}}_{\bm{k}'}^{(R)}(t)
	\end{split}	
	\end{equation}
	\begin{equation} \label{eq:p3_3}
	\begin{split}
	\dot{\alpha}_{\bm{k}'}^{(I)}(t) = &
	(2C_{\bm{k}})^{-1}\sum_{\bm{k}\in\Omega_1} \alpha_{\bm{k}}^{(R)}(t)  \Psi_{3,7,11}(\bm{k},\bm{k}')+  C_{\bm{k}}^{-1}\sum_{\bm{k}\in\Omega_2} \alpha_{\bm{k}}^{(R)}(t)  \Psi_{3,7,11}(\bm{k},\bm{k}') \\ & +  C_{\bm{k}}^{-1}\sum_{\bm{k}\in\Omega_2} \alpha_{\bm{k}}^{(I)}(t)  \Psi_{4,8,12}(\bm{k},\bm{k}') + \beta_{\bm{k}'}^{(I)}(t) + \tilde{\bm{\varepsilon}}_{\bm{k}'}^{(I)}(t)
	\end{split}	
	\end{equation}
	\textit{where} $\Psi_{1,5,9}=\Psi_1+\Psi_5+\Psi_9$, $\Psi_{2,6,10}=\Psi_2+\Psi_6+\Psi_{10}$, $\Psi_{3,7,11}=\Psi_3+\Psi_7+\Psi_{11}$, $\Psi_{4,8,12}=\Psi_4+\Psi_8+\Psi_{12}$, 
	\textit{and}
	
	$\Psi_1(\bm{k},\bm{k}') = \int\bm{\vec{v}}_{\bm{s}}^T\tilde{\bm{k}}f_{\bm{k}}^{(I)}(\bm{s})f_{\bm{k}'}^{(R)}(\bm{s}) d\bm{s}$,
	$\Psi_2(\bm{k},\bm{k}') = -\int\bm{\vec{v}}_{\bm{s}}^T\tilde{\bm{k}}f_{\bm{k}}^{(R)}(\bm{s})f_{\bm{k}'}^{(R)}(\bm{s}) d\bm{s}$, 
	
	$\Psi_3(\bm{k},\bm{k}') = \int\bm{\vec{v}}_{\bm{s}}^T\tilde{\bm{k}}f_{\bm{k}}^{(I)}(\bm{s})f_{\bm{k}'}^{(I)}(\bm{s}) d\bm{s}$, 
	$\Psi_4(\bm{k},\bm{k}') = -\int\bm{\vec{v}}_{\bm{s}}^T\tilde{\bm{k}}f_{\bm{k}}^{(R)}(\bm{s})f_{\bm{k}'}^{(I)}(\bm{s}) d\bm{s}$,
	
	 $\Psi_5(\bm{k},\bm{k}') = \int(-\tilde{\bm{k}}^T \bm{D}_{\bm{s}} \tilde{\bm{k}} f_{\bm{k}}^{(R)} - [\triangledown\cdot \bm{D}_{\bm{s}}]^T \tilde{\bm{k}} f_{\bm{k}}^{(I)} ) f_{\bm{k}'}^{(R)} d\bm{s}$, 
	 
	 $\Psi_6(\bm{k},\bm{k}') = \int(-\tilde{\bm{k}}^T \bm{D}_{\bm{s}} \tilde{\bm{k}} f_{\bm{k}}^{(I)} - [\triangledown\cdot \bm{D}_{\bm{s}}]^T \tilde{\bm{k}} f_{\bm{k}}^{(R)} ) f_{\bm{k}'}^{(R)} d\bm{s}$,
	
	$\Psi_7(\bm{k},\bm{k}') = \int(-\tilde{\bm{k}}^T \bm{D}_{\bm{s}} \tilde{\bm{k}} f_{\bm{k}}^{(R)} - [\triangledown\cdot \bm{D}_{\bm{s}}]^T \tilde{\bm{k}} f_{\bm{k}}^{(I)} ) f_{\bm{k}'}^{(I)} d\bm{s}$,

	$\Psi_8(\bm{k},\bm{k}') = \int(-\tilde{\bm{k}}^T \bm{D}_{\bm{s}} \tilde{\bm{k}} f_{\bm{k}}^{(I)} - [\triangledown\cdot \bm{D}_{\bm{s}}]^T \tilde{\bm{k}} f_{\bm{k}}^{(R)} ) f_{\bm{k}'}^{(I)} d\bm{s}$,

	$\Psi_9(\bm{k},\bm{k}') = -\int \zeta_{\bm{s}} f_{\bm{k}}^{(R)}f_{\bm{k}'}^{(R)}d\bm{s}$,
	$\Psi_{10}(\bm{k},\bm{k}') = -\int\zeta_{\bm{s}} f_{\bm{k}}^{(I)}f_{\bm{k}'}^{(R)}d\bm{s}$,
	$\Psi_{11}(\bm{k},\bm{k}') = -\int \zeta_{\bm{s}} f_{\bm{k}}^{(R)}f_{\bm{k}'}^{(I)}d\bm{s}$,
	
	$\Psi_{12}(\bm{k},\bm{k}') = -\int \zeta_{\bm{s}} f_{\bm{k}}^{(I)}f_{\tilde{\bm{k}}}^{(I)}d\bm{s}$,

	$C_{\bm{k}}=\int f_{\bm{k}}^{(R)}f_{\bm{k}}^{(R)}d\bm{s}=\int f_{\bm{k}}^{(I)}f_{\bm{k}}^{(I)}d\bm{s}$, and $\bm{\tilde{k}}=2\pi\bm{k}$.
\end{proposition1_3*}
\vspace{0.1in}

\textit{\textbf{Remark 1}} (\textit{nonlinear energy transfer between multiple scales}). The proof of Proposition 3 can be obtained using the Galerkin method and is provided in Appendix C.
In Proposition 3, $\Psi_{1}, ..., \Psi_{4}$ determine the transition of the Fourier mode $\alpha^{(R)}_{\bm{k}}(t)-\imath \alpha^{(I)}_{\bm{k}}(t)$ due to convection under the velocity field $\bm{\vec{v}}_{\bm{s}}$. In a special case when $\bm{\vec{v}}_{\bm{s}}$ is constant while the diffusion and source-sink are ignored, the magnitude (i.e., ``energy'') of each Fourier mode is conserved under $\Psi_{1}, ..., \Psi_{4}$ only with shifted argument. This can be clearly shown by letting $\bm{\vec{v}}_{\bm{s}}=\bm{\vec{v}}$, $\bm{D}_{\bm{s}}=\bm{0}$, $\beta_{\bm{k}}^{(R)}=\beta_{\bm{k}}^{(I)}=\tilde{\bm{\varepsilon}}_{\bm{k}'}^{(R)}(t)=\tilde{\bm{\varepsilon}}_{\bm{k}'}^{(I)}(t)=0$. Then, for $\bm{k}' \in \Omega_2$, we have $\Psi_{1}=\Psi_{4}=0$ and $\Psi_{2} = -\Psi_{3} = -2C_{\bm{k}'}\pi\bm{\vec{v}}\bm{k}'$. It follows from (\ref{eq:p3_2}) and (\ref{eq:p3_3}) that
\begin{equation} 
\dot{\alpha}_{\bm{k}'}^{(R)}(t) = -2C_{\bm{k}'}\pi\alpha_{\bm{k}'}^{(I)}(t)\bm{\vec{v}}\bm{k}', \quad\quad \dot{\alpha}_{\bm{k}'}^{(I)}(t) = 2C_{\bm{k}'}\pi\alpha_{\bm{k}'}^{(R)}(t)\bm{\vec{v}}\bm{k}'.
\end{equation}

Hence, 
\begin{equation} 
\begin{split}
\frac{d}{dt}\left\{(\alpha_{\bm{k}'}^{(R)}(t))^2 + (\alpha_{\bm{k}'}^{(I)}(t))^2\right\} & =  2\alpha_{\bm{k}'}^{(R)}(t)\dot{\alpha}_{\bm{k}'}^{(R)}(t) + 2\alpha_{\bm{k}'}^{(I)}(t)\dot{\alpha}_{\bm{k}'}^{(I)}(t) \\
& = 4C_{\bm{k}'}\pi\bm{\vec{v}}\bm{k}'(-\alpha_{\bm{k}'}^{(I)}(t)\alpha_{\bm{k}'}^{(R)}(t) + \alpha_{\bm{k}'}^{(I)}(t)\alpha_{\bm{k}'}^{(R)}(t))= 0
\end{split}
\end{equation}
which implies the conservation of energy at each Fourier mode $\alpha^{(R)}_{\bm{k}}(t)-\imath\alpha^{(I)}_{\bm{k}}(t)$ under the special case with constant convection, and zero diffusion and source-sink. 

However, $\bm{\vec{v}}_{\bm{s}}$ varies in space in our paper, and the evolution of multiple Fourier modes must interact with each other as clearly shown in (\ref{eq:p3_1})$\sim$(\ref{eq:p3_3}) (also see the remark under Proposition 1). Hence, although the convection under a spatially-varying velocity field $\bm{\vec{v}}_{\bm{s}}$ does not increase nor decrease the total energy within the system, it re-distributes the energy to different Fourier modes causing the non-linear energy transfer between multiple scales.

\textit{\textbf{Remark 2}} (\textit{The IDE-based representation}). One important modeling approach to capture the spatio-temporal dynamics relies on the stochastic Integro-Difference Equation (IDE) models \citep{Wikle1999, Wikle2002, Brown2000, Liu2016}. Assuming discrete time steps with lag $\Delta$, the key idea of the IDE-based model is to describe the spatio-temporal dynamics through integral operations in space:
\begin{eqnarray} \label{eq:IDE}
\xi(t,\bm{s}) = \lambda_{\Delta}\int_{\mathbb{S}}\omega_{\bm{s}}(\bm{x})\xi(t-\Delta,\bm{x})d\bm{x} + \text{noise}
\end{eqnarray}
where $\lambda_{\Delta}$ is a scaling factor and $\omega$ is a redistribution kernel. For example, if (\ref{eq:IDE}) is a convolution operation, then, $\omega$ is the convolution kernel. 

\textit{Under spatially and temporally invariant convection-diffusion}, the link between an IDE-based model (in its limiting case) and the convection-diffusion equation has been established \citep{Brown2000, Sigrist2015}. \textit{Under spatially and temporally varying convection-diffusion}, \cite{Liu2016} adopted the IDE-based model without explicitly establishing the connection between the IDE-based model and convection-diffusion equations. 

The following proposition provides the (discretized) IDE-based representation of the SPDE (\ref{eq:SPDE1}) with a \textit{spatially-varying} convection-diffusion operator $\mathcal{A}_{\bm{s}}$. As shown by this proposition, the IDE-based representation can no longer be seen as a spatial convolution operation due to the nonlinear energy transfer between multiple scales discussed in Remark 1 above. 

\begin{proposition1_4*}
For a spatially-varying but temporally-invariant convection-diffusion operator $\mathcal{A}$, the convection-diffusion operation $\mathcal{A}\tilde{\xi}(t+\Delta,\bm{s})$ on the approximated process $\tilde{\xi}(t+\Delta,\bm{s})$ admits a discretized IDE-based representation:
\begin{eqnarray} \label{eq:IDE2}
\tilde{\xi}(t+\Delta,\bm{s}) =\frac{1}{N} \sum_{i=1}^{N} \omega_{\bm{s}}(\bm{x}_i) \tilde{\xi}(t,\bm{x}_i)
\end{eqnarray}
where $N=N_1 \times N_2$, $\bm{s}\in \mathbb{S}$, $\bm{x}_i \in \mathbb{S}$ for $i=1,...,N$, and 
\begin{eqnarray} 
\omega_{\bm{s}}(\bm{x}_i) = \sum_{j}^{N}\sum_{j'}^{N}\left[e^{\bm{G}\Delta}\right]_{j,j'}e^{\imath (\bm{k}_j\bm{s} - \bm{k}_{j'}\bm{x}_i)}.
\end{eqnarray}
\end{proposition1_4*}

The proof of Proposition 4 is given in the Appendix D. Note that:
\begin{itemize}
	\item Although $\omega_{\bm{s}}(\bm{x}_i)$ in (\ref{eq:IDE2}) can still be interpreted as the re-distribution kernel, (\ref{eq:IDE2}) is \textit{no longer a convolution operation} as is often the case in the literature. This is  due to the interesting non-linear energy re-distribution across different frequencies. The matrix exponential, $e^{\bm{G}\Delta}$, is non-diagonal when $\mathcal{A}$ varies in space, and the temporal evolution of the components in $\bm{\alpha}(t)$ is coupled. 
	\item In a special case when the convection-diffusion operator $\mathcal{A}$ does not vary in space and time, the discrete IDE-based representation in Proposition 4 reduces to a convolution operation. Note that when $\mathcal{A}$ does not vary in space and time, $\bm{G}$ becomes a diagonal matrix and $\omega_{\bm{s}}(\bm{x}_i) = \sum_{j=1}^{N}\mathcal{G}_{j}e^{\imath \bm{k}_j(\bm{s} - \bm{x}_i)}$ with $\mathcal{G}_j = -\bm{k}^T_j \bm{D}\bm{k}_j - \zeta - \imath \bm{\vec{v}}^T \bm{k}_j$ being the $i$th diagonal component. Hence, (\ref{eq:IDE2}) reduces to
	\begin{eqnarray}  \label{eq:omega}
	\begin{split}
	\tilde{\xi}(t+\Delta,\bm{s})  & = \frac{1}{N}\sum_{i=1}^{N} \left[\sum_{j=1}^{N}\mathcal{G}_{j}e^{\imath \bm{k}_j(\bm{s} - \bm{x}_i)}\right]\tilde{\xi}(t,\bm{x}_i) \\
	& = \frac{1}{N}\sum_{i=1}^{N} \omega_{\bm{s}}(\bm{x}_i)\tilde{\xi}(t,\bm{x}_i).
	\end{split}
	\end{eqnarray}
	In such a special case, the dynamics of $\tilde{\xi}(t,\bm{s})$ is captured by spatial convolution and is consistent with the existing results in the literature \citep{Wikle1999, Brown2000, Liu2016}. In fact, it is easy to see that $\omega_{\bm{s}}(\bm{x}_i)$ is the inverse Fourier transform of $\mathcal{G}_j$, and $\omega_{\bm{s}}(\bm{x}_i)$ is a Gaussian kernel (Section 3.2.1 in \cite{Liu2016}).  
	Of course, in a general case where $\mathcal{A}$ varies in space, the dynamics of $\tilde{\xi}(t,\bm{s})$ is captured by ($\ref{eq:IDE2}$), which is no longer a convolution operation in space. 
\end{itemize}

Proposition 4 links the proposed model to a large body of literature and provides a different perspective on how the proposed model can be interpreted.

\section{Integration with a Dynamical Model} \label{sec:DSTM}
The results obtained in Sections \ref{sec:general} and \ref{sec:special} enable us to embed the physics, governed by the convection-diffusion process (\ref{eq:SPDE1}), into the framework of Hierarchical Spatio-Temporal Models (DSTM) described in \cite{Wikle1998, Cressie2011, Katzfuss2019}. The DSTM has been shown to be effective in handling non-stationary, irregular complex processes with either continuous, discrete or multivariate data, and provides computational advantages for inference, monitoring, prediction and dynamic network design.
The DSTM framework consists of a data model, a process model, and a parameter model. The data model can be directly obtained from the spectral decomposition (\ref{eq:decomposition}) and is given by:
\begin{equation} \label{eq:data_model}
\textrm{Data Model: } \bm{Y}(t) = \bm{F} \bm{\alpha}(t) + \bm{V}(t)
\end{equation}
where $\bm{Y}(t) = (Y_{\bm{s}_1}(t),...,Y_{\bm{s}_N}(t))^T$ is a $N \times 1$ vector that contains the data observed at time $t$, $\bm{\alpha}(t)$ 
is a $K \times 1$ vector that contains the spectral coefficients, $\bm{F} =(\bm{f}(\bm{s}_1),\bm{f}(\bm{s}_2),...,\bm{f}(\bm{s}_N))^T$ is a $N \times K$ observation matrix of Fourier functions, and the noise term $\bm{V}(t)$ is added to capture the unexplained variation such as observation error. 

The process model, which captures the temporal dynamics of $\bm{\alpha}(t)$, can be obtained from the discrete-time representation of the ODE (\ref{eq:gamma_transition_2}): 
\begin{equation} \label{eq:process_model}
\textrm{Process Model: } \bm{\alpha}(t) = \exp(\bm{G}\Delta) \bm{\alpha}(t-\Delta) + \bm{\beta}(t-\Delta)\Delta + \bm{W}_{\bm{\alpha}}(t)
\end{equation}
where $\bm{W}_{\bm{\alpha}}(t)\sim N(0,\int_{0}^{\Delta}\exp(\bm{G}(\Delta-\tau))\bm{H}\exp^*(\bm{G}^T(\Delta-\tau))d\tau)$. 
 
If necessary, a parameter model 
can also be included to incorporate prior information or regularizations obtained from other sources, especially the outputs from physical models which are widely available for many spatio-temporal modeling problems. For example, when modeling the temperature field in a data center, information on the velocity field (i.e., air flow) and diffusion matrix can usually be derived from the CFD model and flow sensors. Another example is the modeling of urban air pollution process or the motion of extreme weather systems, information on wind field is usually available from the Numerical Weather Prediction (NWP) models \citep{Liu2016, Liu2018, Lu2017}. 

In the convection-diffusion process (\ref{eq:SPDE1}), $Q(t,\bm{s})$ is the deterministic source (i.e., generation) and sink (i.e., destruction) of the quantify of interest. 
As discussed in Section \ref{sec:general}, $Q(t,\bm{s})$ admits a spectral representation $Q(t,\bm{s})  = \bm{f}^T(\bm{s})\bm{\beta}(t)$ with $\bm{\beta}(t) = (\beta^{(1)}(t),...,\beta^{(K)}(t) )^T$ dynamically evolving over time.
If an AR(1) process is considered for $\bm{\beta}(t)$, we obtain

\vspace{0.1in}
\begin{proposition1_5*}
\textit{Under the assumptions A1$\sim$A4 made in Proposition 1 and consider an AR(1) process for $\bm{\beta}(t)$ such that $\bm{\beta}(t) = \bm{M}_\Delta\bm{\beta}(t-\Delta)+\bm{W}_{\bm{\beta}}(t)$ where $\bm{W}_{\bm{\beta}}(t)\sim N(0, \tau_{\bm{\beta}}^2\bm{I}_{K})$
	and $\bm{M}_\Delta=\rho_\Delta\bm{I}_{K}$, a dynamical model with local linear growth can be written as:
}
	\begin{eqnarray} \label{eq:dynamical}
	\begin{split}
	& \bm{Y}(t) = (\bm{F} , \bm{0}_{N, K})\bm{\theta}(t)+ \bm{V}(t) \\
	& \bm{\theta}(t) = \bm{\tilde{G}}\bm{\theta}(t-\Delta)+ \bm{W}(t)
	\end{split}
	\end{eqnarray}
	\textit{where }
	\begin{eqnarray} \label{eq:tildeG}
	\bm{\tilde{G}} = \left(\begin{array}{c|c}
	\exp(\bm{G}\Delta) & \Delta\bm{I}_{K}  \\ \hline
	\bm{0}_{K} & \bm{M}_\Delta
	\end{array}\right)
	\end{eqnarray}
\textit{and} $\bm{\theta}(t)=(\bm{\alpha}(t),\bm{\beta}(t))^T$, 
$\bm{V}(t) \sim N(0, \tau_{\bm{V}}^2\bm{I}_{N})$,  $\bm{W}(t)=(\bm{W}_{\bm{\alpha}}(t), \bm{W}_{\bm{\beta}}(t))^T$. 
\end{proposition1_5*}
\vspace{0.1in}

It is easy to see that $\bm{W}(t) \sim N(0, \bm{\Sigma}_{\bm{W}})$, and
\begin{eqnarray}
\bm{\Sigma}_{\bm{W}}= \left(\begin{array}{c|c}
\int_{0}^{\Delta}\exp(\bm{G}(\Delta-\tau))\bm{H}\exp^*(\bm{G}^T(\Delta-\tau))d\tau &  \\ \hline
 & \tau_{\bm{\beta}}^2\bm{I}_{K}
\end{array}\right).
\end{eqnarray}

For large spatio-temporal data, the traditional geostatistical modeling paradigm is limited due to the prohibitive $\mathcal{O}((TN)^3)$ operations associated with factorizing the dense covariance matrices.
The conditioning structure of the proposed DSTM reduces the cubic time complexity to $\mathcal{O}(TN^3)$ if the Kalman filter and FFBS (Forward Filtering and Backward Sampling) are used \citep{Carter1994}. 
More importantly, since the data process (\ref{eq:data_model}) is based on the spectral representation of a spatial process, \cite{Sigrist2015} noted that the Kalman filter can be efficiently performed in the spectral space for gridded data by converting $\bm{Y}(t)$ from the physical space to $\bm{\tilde{Y}}(t)$ in the spectral space using FFT which requires $\mathcal{O}(TN\log(N))$.
Hence, the dynamical model (\ref{eq:dynamical}) in the spectral space is given by
\begin{eqnarray} \label{eq:dynamical_spectral}
\begin{split}
& \bm{\tilde{Y}}(t) = \bm{\tilde{F}}\bm{\theta}(t)+ \bm{\tilde{V}}(t) \\
& \bm{\theta}(t) = \bm{\tilde{G}}\bm{\theta}(t-\Delta)+ \bm{W}(t)
\end{split}
\end{eqnarray}
where $\bm{\tilde{Y}}(t)$ is the Fourier transform of $\bm{Y}(t)$ obtained from FFT, $\bm{\tilde{F}}=(\bm{I}_K , \bm{0}_K)$, and $\bm{\tilde{V}}(t)  \sim N(0, \tau_{\bm{\tilde{V}}}^2\bm{I}_{K})$. 

Note that the dimensions of $\bm{\tilde{Y}}(t)$, $\bm{\theta}(t)$ and $\bm{\tilde{F}}$ are respectively $K \times 1$, $2K \times 1$ and $K \times 2K$. In practice, it is possible to only keep the low frequency components and make $K$ much smaller than $N$ (see Figure \ref{fig:fig2}). Therefore, the choice of $K$ in (\ref{eq:decomposition}) controls both the dimension and computational cost of the problem. 

The  Kalman Filter in the spectral domain can be obtained from the classical iterations \citep{West1997, Petris2009}.
Consider a dynamical model specified by (\ref{eq:dynamical_spectral}). Let $\bm{m}_{t|t-\Delta}$ and $\bm{m}_{t|t}$ respectively be the one-step-ahead predictive and filtering means of $\bm{\theta}(t)$, let $\bm{Q}_{t|t-\Delta}$ and $\bm{Q}_{t|t}$ respectively be the one-step-ahead predictive and filtering covariance matrices of $\bm{\theta}(t)$, and let $\bm{\theta}(0)\sim \mathcal{N}(\bm{m}_{0|0}, \bm{Q}_{0|0})$, then, the following statements hold:

The one-step-ahead predictive mean and covariance matrix of $\bm{\theta}(t)$ are
\begin{eqnarray} \label{eq:predictive}
\begin{split}
& \bm{m}_{t|t-\Delta} =  \bm{\tilde{G}}\bm{m}_{t-\Delta|t-\Delta} \\
& \bm{Q}_{t|t-\Delta} = \bm{\tilde{G}}\bm{Q}_{t-\Delta|t-\Delta}\bm{\tilde{G}}^T + \bm{\Sigma}_{\bm{W}},
\end{split}
\end{eqnarray}
and the filtering mean and covariance matrix of $\bm{\theta}(t)$ are
\begin{eqnarray} \label{eq:filtering}
\begin{split}
& \bm{m}_{t|t} =  \bm{\tilde{G}}\bm{m}_{t|t-\Delta}+ 
\bm{Q}_{t|t-\Delta}^{[1:2K,1:K]}  
 (\bm{Q}_{t|t-\Delta}^{[1:K,1:K]}+\tau_{\bm{\tilde{V}}}\bm{I}_K)^{-1}(\tilde{\bm{Y}}(t)-\bm{m}_{t|t-\Delta}^{[1:K,1]}) \\
& \bm{Q}_{t|t} =  \bm{Q}_{t|t-\Delta} -\bm{Q}_{t|t-\Delta}^{[1:2K,1:K]} (\bm{Q}_{t|t-\Delta}^{[1:K,1:K]}+\tau_{\bm{\tilde{V}}}\bm{I}_K)^{-1} \bm{Q}_{t|t-\Delta}^{[1:K,1:2K]}.
\end{split}
\end{eqnarray}
Here, $\cdot^{[1:K_1,1:K_2]}$ returns the first $K_1$ rows and $K_2$ columns of a matrix,  $\bm{\tilde{F}}\bm{Q}_{t|t-\Delta}\bm{\tilde{F}}^T=\bm{Q}_{t|t-\Delta}^{[1:K,1:K]}$, $\bm{Q}_{t|t-\Delta}\bm{\tilde{F}}^T=\bm{Q}_{t|t-\Delta}^{[1:2K,1:K]}$, $\bm{\tilde{F}}\bm{Q}_{t|t-\Delta}=\bm{Q}_{t|t-\Delta}^{[1:K,1:2K]}$ and $\bm{\tilde{F}}\bm{m}_{t|t-\Delta}=\bm{m}_{t|t-\Delta}^{[1:K,1]}$. 
 

The dynamical models (\ref{eq:dynamical}) and (\ref{eq:dynamical_spectral}) can be extended to non-Gaussian models, and the Sequential Monte Carlo (SMC) method can then be used; see Appendix E for a GPU-accelerated SMC algorithm.  

The dynamical model (\ref{eq:dynamical_spectral}) contains a large number of unknown parameters: (i) the velocity field $\bm{\vec{v}}_{\bm{s}}$, diffusivity $\bm{D}_{\bm{s}}$, and decay $\zeta_{\bm{s}}$ which vary in space; (ii) $\rho_\Delta$ and $\tau_{\bm{\beta}}$ defined above (\ref{eq:dynamical}); (iii) the spectral density $\bm{H}=\mathrm{diag}(\tilde{h}(\bm{k}_j))$ of $\bm{\tilde{\varepsilon}}(t)$; and (iv) $\tau_{\bm{\tilde{V}}}$. 
Here, $\bm{\vec{v}}_{\bm{s}}$, $\bm{D}_{\bm{s}}$, $\zeta_{\bm{s}}$ and $\rho_\Delta$ are needed for computing $\tilde{\bm{G}}$ in (\ref{eq:tildeG}), $\rho_\Delta$ and $\bm{H}$ are needed for evaluating the covariance matrix $\bm{\Sigma}_{\bm{W}}$, and $\tau_{\bm{\tilde{V}}}$ determines the covariance matrix of $\bm{\tilde{V}}$.

The log-likelihood function can be constructed based on the one-step-ahead predictive distribution of $\bm{\tilde{Y}}(t)$ obtained from the Kalman Filter iterations:
\begin{eqnarray} \label{eq:likelihood}
\begin{split}
l = & -\frac{1}{2}\sum_{t=1}^{T}\log \det(\bm{Q}_{t|t-\Delta}^{[1:K,1:K]}+\tau_{\bm{\tilde{V}}}\bm{I}_K) \\ & -\frac{1}{2}(\tilde{Y}(t)-\bm{m}_{t|t-\Delta}^{[1:K]})^T     (\bm{Q}_{t|t-\Delta}^{[1:K,1:K]}+\tau_{\bm{\tilde{V}}}\bm{I}_K)^{-1}(\tilde{Y}(t)-\bm{m}_{t|t-\Delta}^{[1:K]}). 
\end{split}
\end{eqnarray}

Note that the problem apparently becomes intractable if 
we allow the velocity $\bm{\vec{v}}_{\bm{s}}$, diffusivity $\bm{D}_{\bm{s}}$, and decay $\zeta_{\bm{s}}$ to arbitrarily vary in the spatial domain. Hence, it is necessary to impose some assumptions (such as spatial smoothness) on $\bm{\vec{v}}_{\bm{s}}$, $\bm{D}_{\bm{s}}$ and $\zeta_{\bm{s}}$. In this paper, we consider a flexible locally weighted mixture of linear regression models \citep{Stroud2001} for the velocity $\bm{\vec{v}}_{\bm{s}}$, although other modeling approaches are possible depending on the problem of interest. Let $v_{\bm{s}}^{(x)}$ and $v_{\bm{s}}^{(y)}$ respectively be the horizontal and vertical components of the velocity vector $\bm{\vec{v}}_{\bm{s}}$ at $\bm{s}$, and let $v^{\mathrm{(max)}}>0$ be the specified maximum speed magnitude, we assume
\begin{eqnarray}
v_{\bm{s}}^{(x)}  = v^{\mathrm{(max)}} \tanh \left\{ \sum_{j=1}^{J}\pi_{j}(\bm{s})\bm{b}_j^T(\bm{s})\bm{\gamma}_{j}^{(x)} \right\}, 
\quad v_{\bm{s}}^{(y)}  = v^{\mathrm{(max)}} \tanh \left\{ \sum_{j=1}^{J}\pi_{j}(\bm{s})\bm{b}_j^T(\bm{s})\bm{\gamma}_{j}^{(y)} \right\} 
\label{eq:mixture}
\end{eqnarray}
where $\bm{b}_j(\bm{s})$ is a column vector of known basis functions, $\bm{\gamma}_{j}^{(x)}$ and $\bm{\gamma}_{j}^{(y)}$ are column vectors of unknown parameters, and $\pi_j$ is a non-negative kernel centered at chosen locations. Note that, the hyperbolic tangent, $\tanh(\cdot)$, is used to bound both $|v_{\bm{s}}^{(x)}|$ and $|v_{\bm{s}}^{(y)}|$ within $[-v^{\mathrm{(max)}},v^{\mathrm{(max)}}]$.

Let $\bm{v}^{(x)}=(v_{\bm{s}_1}^{(x)},...,v_{\bm{s}_N}^{(x)})^T$, $\bm{v}^{(y)}=(v_{\bm{s}_1}^{(y)},...,v_{\bm{s}_N}^{(y)})^T$,
$\bm{B}_{j}=(\bm{b}_j(\bm{s}_{1}), ..., \bm{b}_j(\bm{s}_{N}))$ and $\bm{\pi}_j = (\pi_{j}(\bm{s}_{1}), ..., \pi_{j}(\bm{s}_{N}))$, we have
\begin{eqnarray}
\begin{split}
& \bm{v}^{(x)}  = v^{\mathrm{(max)}} \tanh \left\{ ( \textrm{diag}(\bm{\pi}_1)\bm{B}_{1}, ..., \textrm{diag}(\bm{\pi}_J)\bm{B}_{J})\left(\begin{array}{c}  \bm{\gamma}_{1}^{(x)}\\ \vdots \\ \bm{\gamma}_{J}^{(x)} \end{array}\right) \right\}\\
& \bm{v}^{(y)}  = v^{\mathrm{(max)}} \tanh \left\{ ( \textrm{diag}(\bm{\pi}_1)\bm{B}_{1}, ..., \textrm{diag}(\bm{\pi}_J)\bm{B}_{J})\left(\begin{array}{c}  \bm{\gamma}_{1}^{(y)}\\ \vdots \\ \bm{\gamma}_{J}^{(y)} \end{array}\right) \right\}.
\end{split}
\label{eq:MeanFunction}
\end{eqnarray}

Similarly, it is possible to model the decay $\zeta_{\bm{s}}$ based on the same idea. The diffusivity $\bm{D}_{\bm{s}}$ usually depends on the velocity field $\bm{\vec{v}}_{\bm{s}}$ through some functions motivated by fundamental physics. 
For some applications, prior information about $\bm{\vec{v}}_{\bm{s}}$,  $\bm{D}_{\bm{s}}$, and $\zeta_{\bm{s}}$ is available. For example, in the modeling of weather systems, the velocity, diffusivity and decay can be generated from the Numerical Weather Prediction models. For another example, in the modeling of the temperature field in a DC computer room, the air flow can be obtained from the Navier-Stokes equations. Hence, it is sometimes possible to treat $\bm{\vec{v}}_{\bm{s}}$, $\bm{D}_{\bm{s}}$ and $\zeta_{\bm{s}}$ as known, or, incorporate the prior knowledge on these parameters in a Bayesian framework. 

Compared with existing approaches, such as \cite{Stroud2010}, the proposed approach can be viewed as a more intrusive statistical version of a reduced-order physics model. 
The proposed model is built upon the solution (in the frequency domain) of the
convection-diffusion equations, and the output from our statistical model remains within the solution space of the physical model (i.e., the SPDE (\ref{eq:SPDE1})). The physical model is integrated into a statistical modeling framework with all parameters being estimated from data. A less intrusive approach that integrates the physical model outputs and measurement data are available in \cite{Stroud2010}, which requires some key physical parameters (velocity, diffusion and forcing) to be found before they are integrated into the statistical model. 

\begin{figure}[h!] 
	\begin{center}
		\includegraphics[width=0.95\textwidth]{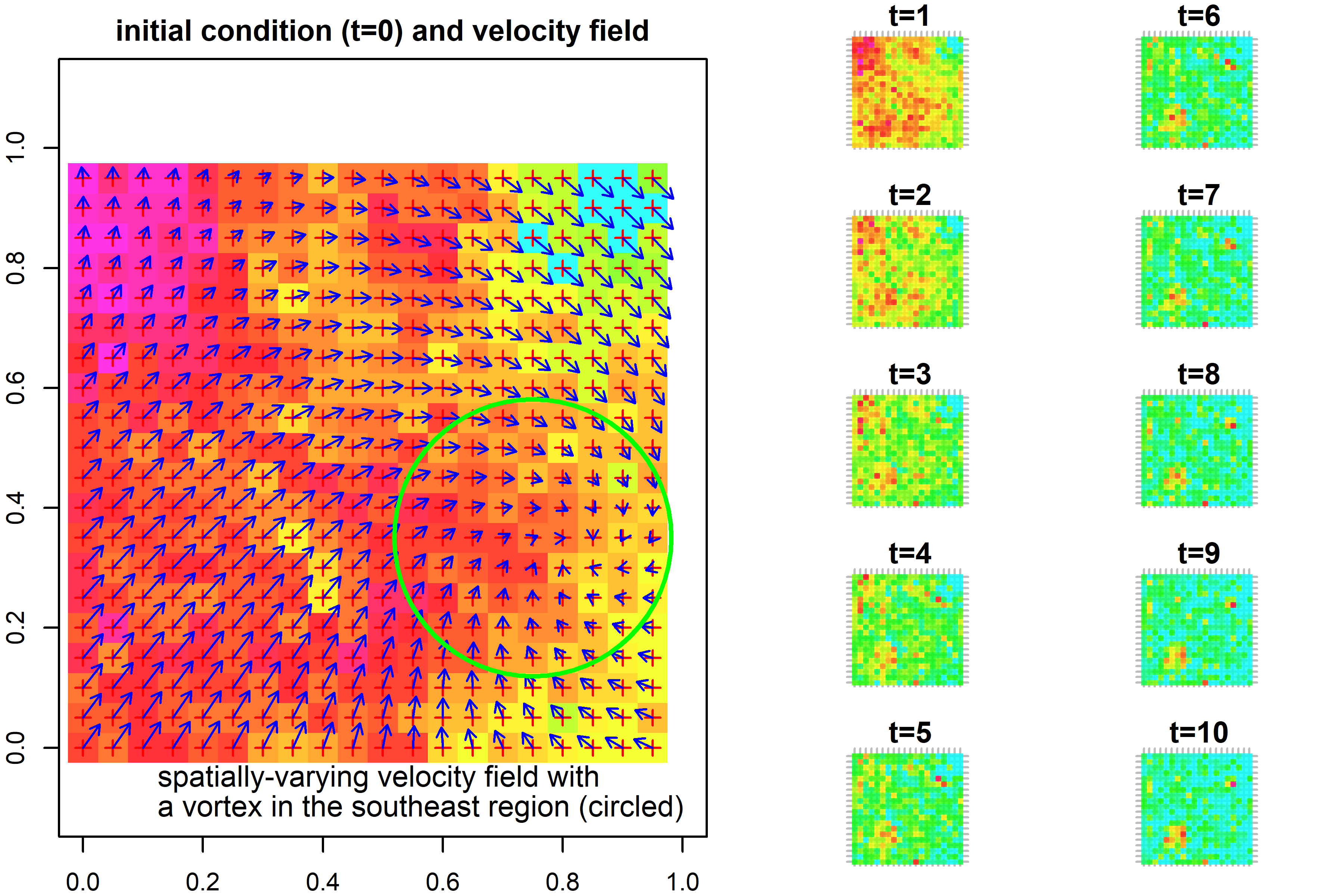}
		\centering
		\caption{The data set (simulated from (\ref{eq:SPDE1})) used for the numerical example. The panel on the left shows the initial condition and a spatially-varying velocity field. The figures on the right show the images for $t=1,2,...,10$.}
		\label{fig:numeric1} 
	\end{center}
\end{figure}

\vspace{-24pt}
\section{Numerical Example and Case Study} \label{sec:numeric}
A numerical example is presented in Section \ref{sec: example} to investigate the basic properties and generate some critical insights of the proposed approach. A case study is followed in Section \ref{sec:case} to illustrate the application of the proposed method on the modeling of weather radar image data for tropical thunderstorms.

\subsection{Numerical Example}\label{sec: example}
The numerical example is based on a simulated spatio-temporal data set shown in Figure \ref{fig:numeric1}. The data are generated from the SPDE (\ref{eq:SPDE1}) on a $20\times 20$ gridded spatial area for $t=0,1,...,10$. Both the initial condition ($t=0$) and velocity field $\bm{\vec{v}}_{\bm{s}}$ are shown on the left panel of Figure \ref{fig:numeric1}. The velocity vectors vary over the spatial domain, and a small \textit{vortex} is created in the southeast region to test the capabilities of the proposed model in capturing such complexities. The diffusivity $\bm{D}_{\bm{s}}$ is set to an identify matrix. The decay $\zeta$ is set to a constant 0.9. The error process $\varepsilon(t,\bm{s})$ is generated by $\bm{f}^T(\bm{s})\tilde{\bm{\varepsilon}}(t)$ where $\tilde{\bm{\varepsilon}}(t)$ is a random
vector with the spectral density $0.05\bm{I}_K$. A spatially-varying source-sink term $Q(\bm{s},t)$ is also included in the process.

The model parameters are estimated by MLE. Here, the velocity field is modeled by (\ref{eq:MeanFunction}) with $J=4$ and $v^{(\mathrm{max})}=0.19$, which is approximately one fifth of the width of the spatial domain in the horizontal or vertical direction. Note that, in the absence of any prior knowledge on the velocity field, the four kernels are uniformly allocated over the spatial domain: $(0.225,0.225)$, $(0.725,0.725)$, $(0.225,0.725)$, $(0.725,0.225)$.  When prior knowledge on the velocity field is available as in some applications, such knowledge can guide us to appropriately place the kernels. In radar-based precipitation nowcasting, for example, information on wind field can be obtained from Numerical Weather Predictions and more kernels can be placed to areas where the variability of wind is expected to be high. 

\begin{figure}[h!] 
	\begin{center}
		\includegraphics[width=0.9\textwidth]{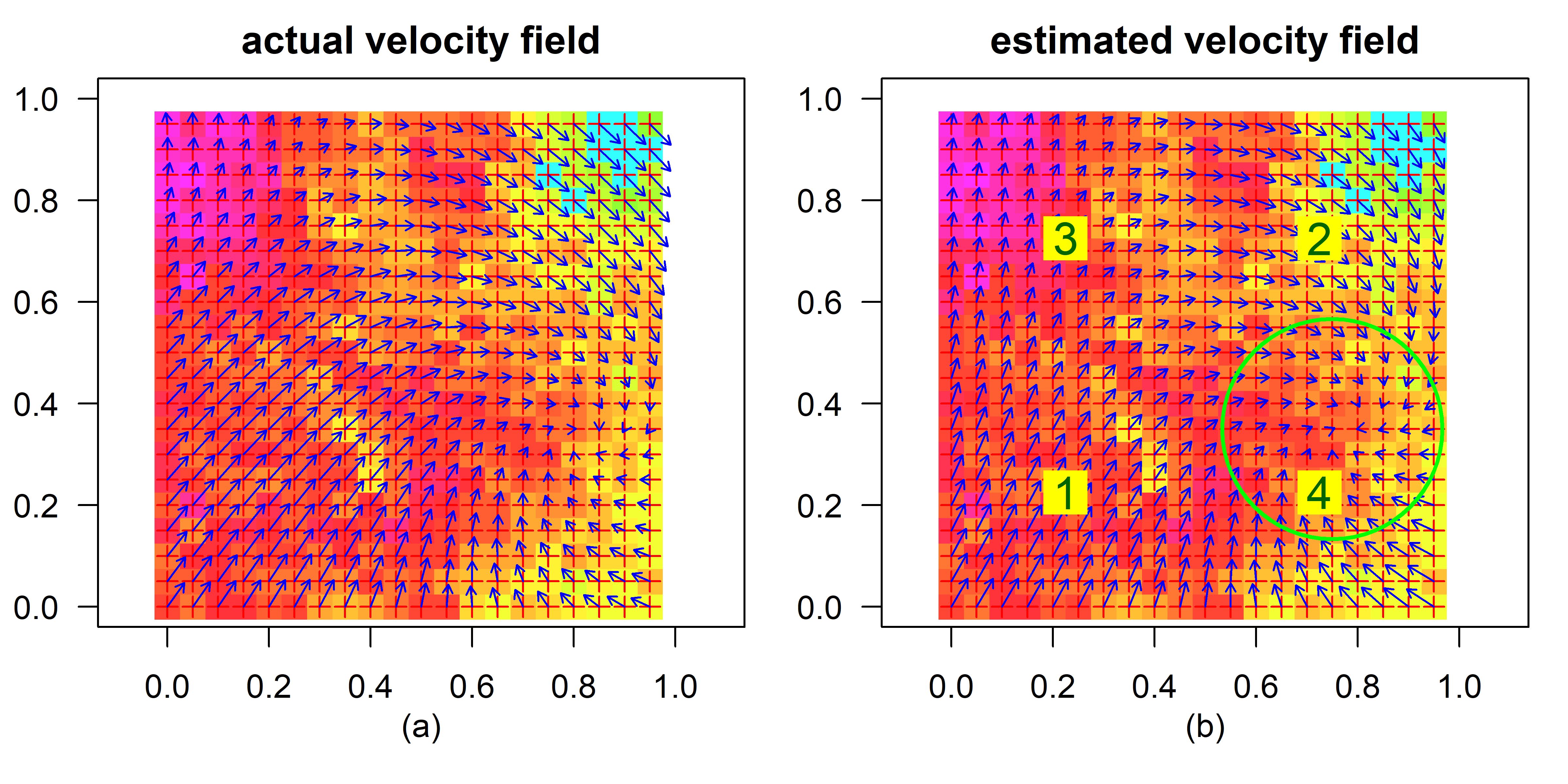}
		\centering
		\caption{Actual (left) and estimated (right) velocity fields}
		\label{fig:numeric2} 
	\end{center}
\end{figure}

\begin{figure}[h!] 
	\begin{center}
		\includegraphics[width=1\textwidth]{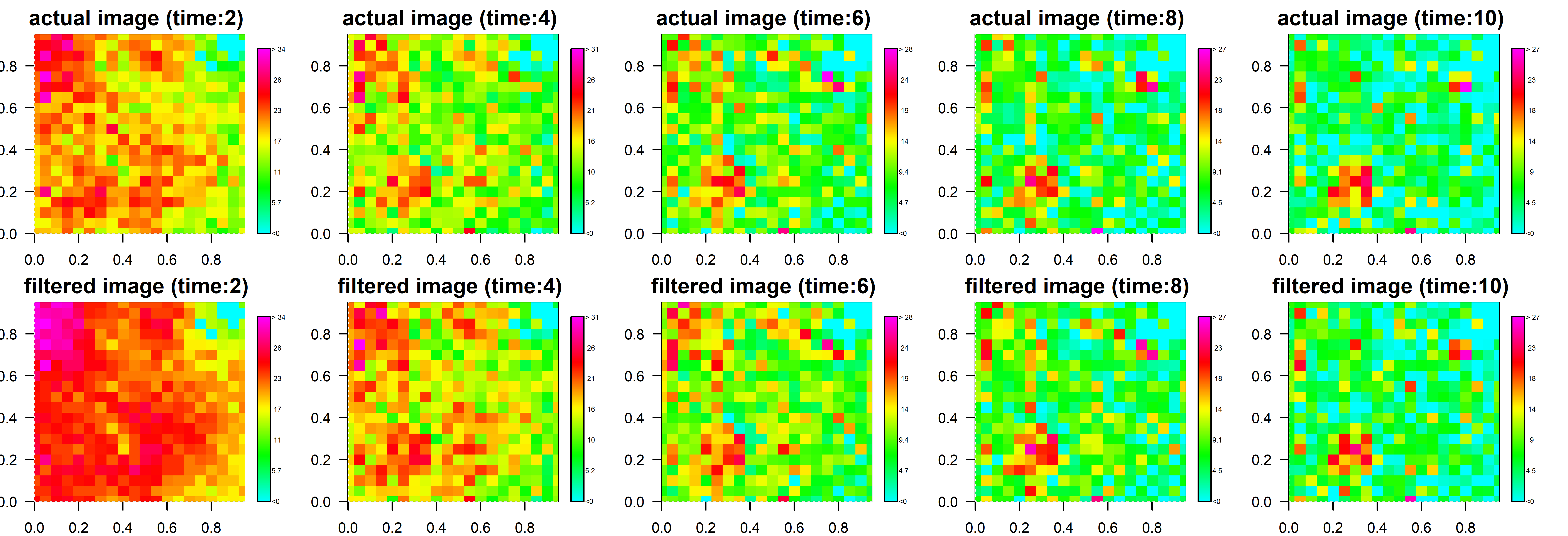}
		\centering
		\caption{Actual (top row) and filtered (bottom row) images at times 2, 4, 6, 8 and 10.}
		\label{fig:numeric3} 
	\end{center}
\end{figure}

Figure \ref{fig:numeric2} shows the actual and estimated velocity fields. 
The centers of the four kernels in (\ref{eq:MeanFunction}) are also shown in the figure. We see that the proposed approach accurately estimates the velocity field which varies in the spatial domain. In particular, the small \textit{vortex} in the southeast region is successfully captured, demonstrating the capabilities of the proposed approach in modeling spatio-temporal data arising from a physical process (\ref{eq:SPDE1}) under a spatially-varying velocity field. The Kalman Filter in spectral domain (Proposition 5) is used to obtained the filtered $\bm{\theta}$, which is a $361\times1$ vector in the dynamical model (\ref{eq:dynamical_spectral}). Then, inverse Fourier transform is applied to construct the filtered images, and the reconstructed images at times $t=2,4,6,8,10$ are shown in Figure \ref{fig:numeric3}.

\subsection{Case Study: Radar-Based Weather System Modeling} \label{sec:case}
Following the numerical example, a case study is presented in which the proposed method is used to model highly chaotic weather systems---tropical thunderstorms---based on weather radar images. 
Weather radar echoes, correlated in both space and time, provide a rich source of information for short-term precipitation nowcasting. In the meteorological community, the methods which are used to perform spatio-temporal extrapolation/advection of radar reflectivity (echo) field are collectively known as the radar-based Quantitative Precipitation Forecasts (QPF) \citep{RMI2008}.

\subsubsection{Basic Settings and Data Processing} \label{sec:data_case}
\begin{figure}[h!] 
	\begin{center}
		\includegraphics[width=1\textwidth]{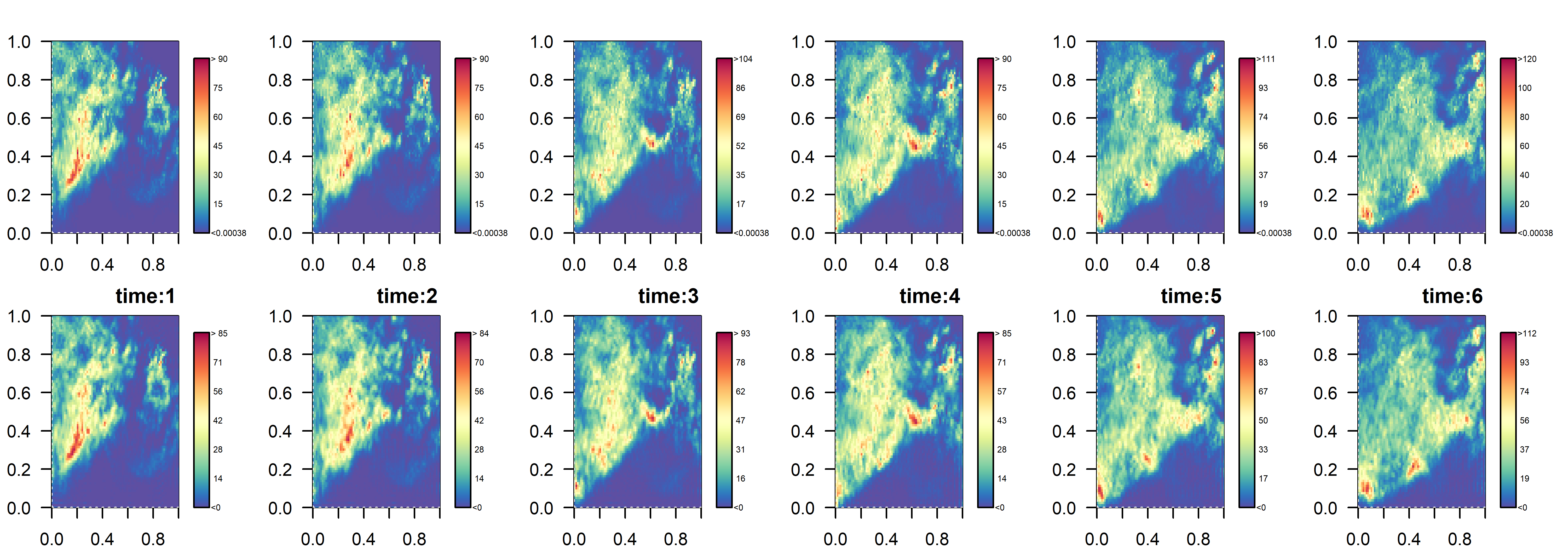}
		\centering
		\caption{A tropical thunderstorm system on a doppler weather radar system at $t=1,2,...,6$. Top row: original CAPPI reflectivity data at 1 km above the mean sea level; Bottom row: low-pass filtered images (threshold: 30 radians per unit distance)}
		\label{fig:radar} 
	\end{center}
\end{figure}

Figure \ref{fig:radar} (the top row) shows a sequence of radar images obtained from a dual polarization Meteorological Doppler Weather Radar (MDWR) system. Each image contains the standard Constant Altitude Plan Position Indicator (CAPPI) reflectivity data at 1 km above the mean sea level. The legend shows the precipitation rate (mm/hr) converted from the CAPPI data using the Marshall-Palmer relationship. It is possible to see that the thunderstorm is moving from west to east driven by a velocity (wind) field, which is not directly observable. 

\begin{figure}[h!] 
	\begin{center}
		\includegraphics[width=0.35\textwidth]{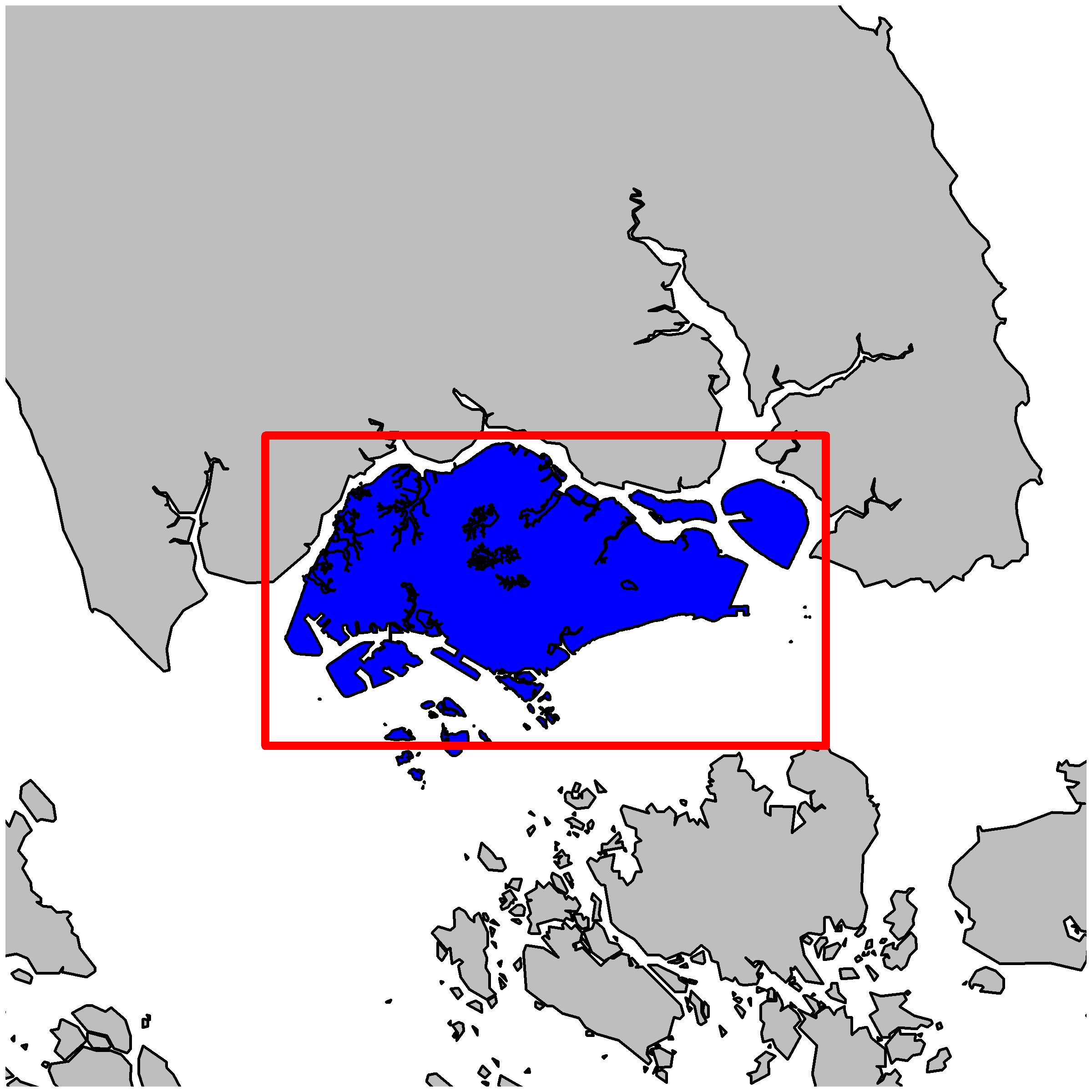}
		\centering
		\caption{The spatial area of interest (covered by the rectangle) and the area covered by a radar image. The dimensions of the area covered by a radar image are approximately 2.5 and 5 times larger than the dimensions of the spatial area of interest, respectively in the horizontal and vertical directions.}
		\label{fig:singapore} 
	\end{center}
\end{figure}

The images are generated at 5-min intervals and on a Cartesian 2D grid of 200$\times$200 pixels. The centers of two neighboring pixels are approximately 0.5km apart, and the spatial area covered by an image is approximately $100\times 100$$\textrm{km}^2$. The radar images are obtained from Singapore---a tropical island city-state which occupies approximately a $40\times 20$$\textrm{km}^2$ spatial area in the center of the image; see Figure \ref{fig:singapore}. Hence, in the horizontal and vertical directions, the dimensions of the area covered by a radar image are respectively 2.5 and 5 times larger than the dimensions of the spatial area of interest. There are three reasons why this setting is necessary: 1) It is known that the 2D DFT suffers from the edge-effects when it is applied to non-periodic images, such as the cross-shaped artifacts in the frequency domain due to spectral leakage. This issue has been extensively discussed in \cite{Guinness2017} and \cite{Guinness2019} from the perspective of statistical modeling; 2) Since the radar-based QPF in the meteorological society essentially relies on the spatio-temporal extrapolation of radar reflectivity field, it is impossible to predict the reflectivity at the (influx) boundaries, where the weather system moves into the range of a radar image; 3) If the entire spatial area covered by a radar image is of interest, then, it is necessary to consider the boundary conditions of the SPDE (\ref{eq:SPDE1}) which are unknown and uncontrollable in this case study. For the three reasons above, the modeling and prediction results are less reliable at the boundaries of the images, and embedding the spatial area of interest (which is much smaller) at the center of the radar image (computational domain) effectively mitigates this boundary issue in practice. 

As tropical storms are highly chaotic with small-scale localized convective cells, radar images at pixel levels are noisy and non-smooth. Hence, we perform FFT on the original images and reconstruct the low-pass filtered images by only retaining the low-frequency components whose (absolute) angular wavenumber is less than or equal 30 radians per unit distance in both directions. The low-pass filtered images, shown in the bottom row of Figure \ref{fig:radar}, are used for estimating the unknown model parameters including the wind field. Note that, the original DSTM framework does not necessarily require such a pre-processing of input images \citep{Cressie2011}. For this particular application, the computational bottleneck is to estimate the spatially-varying wind field which determines the matrix $\bm{G}$. Given the large spatial coverage of a radar image (10,000$\textrm{km}^2$ in this example), the small-scale chaotic local wind patterns are usually not of our main concern. Hence, keeping only the low-frequency terms reduces the computational time and allows us to capture the dominant motion of the weather system over a relatively large spatial domain. Appendix E provides detailed discussions on the computational aspects of the proposed approach. 

\begin{figure}[h!] 
	\begin{center}
		\includegraphics[width=1\textwidth]{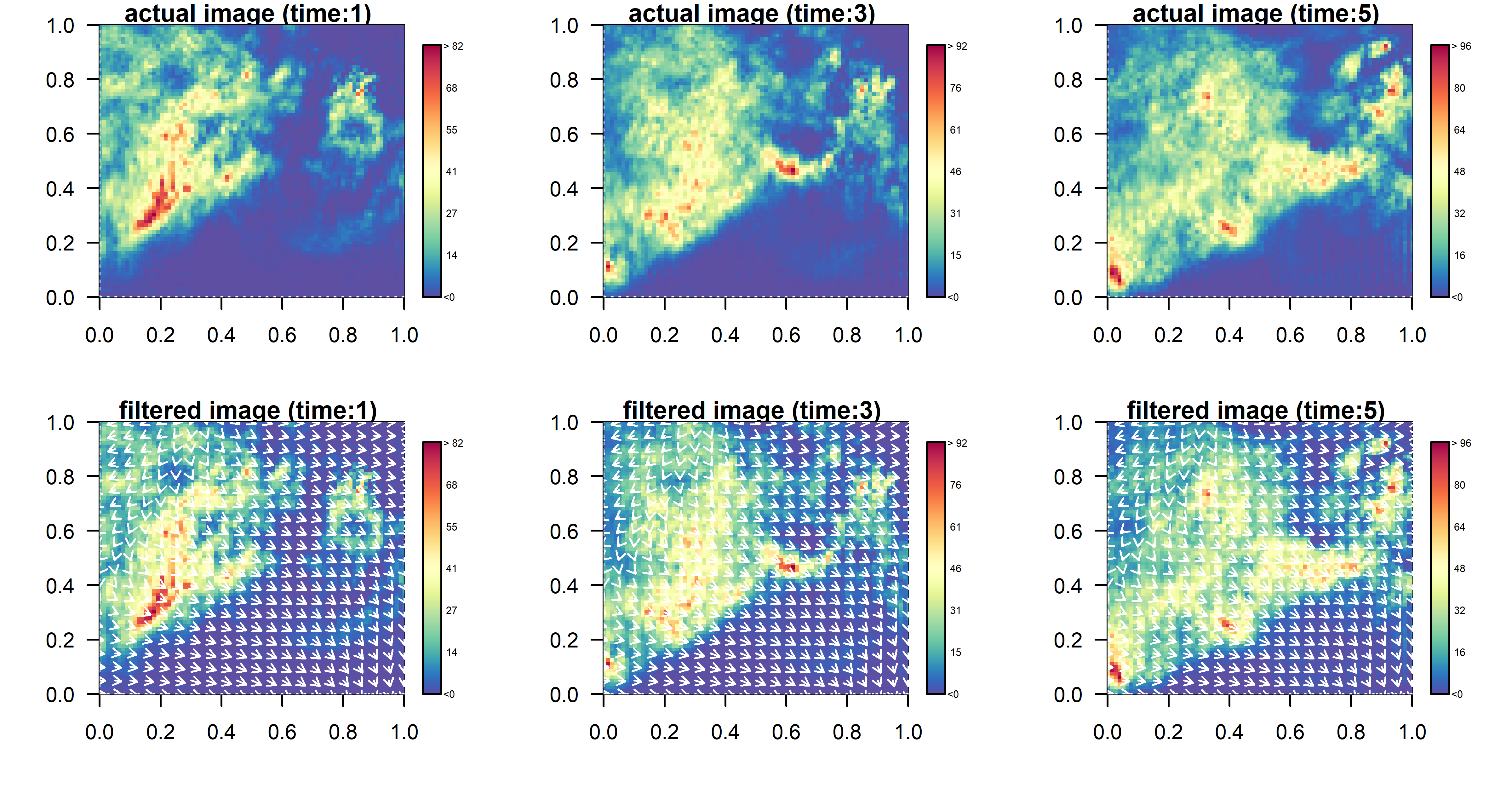}
		\centering
		\caption{Observed and filtered images at selected times, $t=1,3,5$, with the estimated spatially-varying wind field ($\text{km/5 min}$).}
		\label{fig:case_filter} 
	\end{center}
\end{figure}
\subsubsection{Numerical Results}
Figure \ref{fig:case_filter} shows both the observed and filtered images at selected times, $t=1,3,5$. The estimated spatially-varying wind field is also included in the filtered images. It can be seen that the estimated wind field explains the west-to-east motion of the weather system, which is consistent with our initial observation. It is important to note that, for tropical areas where trade winds from both hemispheres meet (known as the Inter-tropical Convergence Zone), the wind fields near the equator are typically highly variable in space and is captured by the proposed approach.

\begin{figure}[h!] 
	\begin{center}
		\includegraphics[width=1\textwidth]{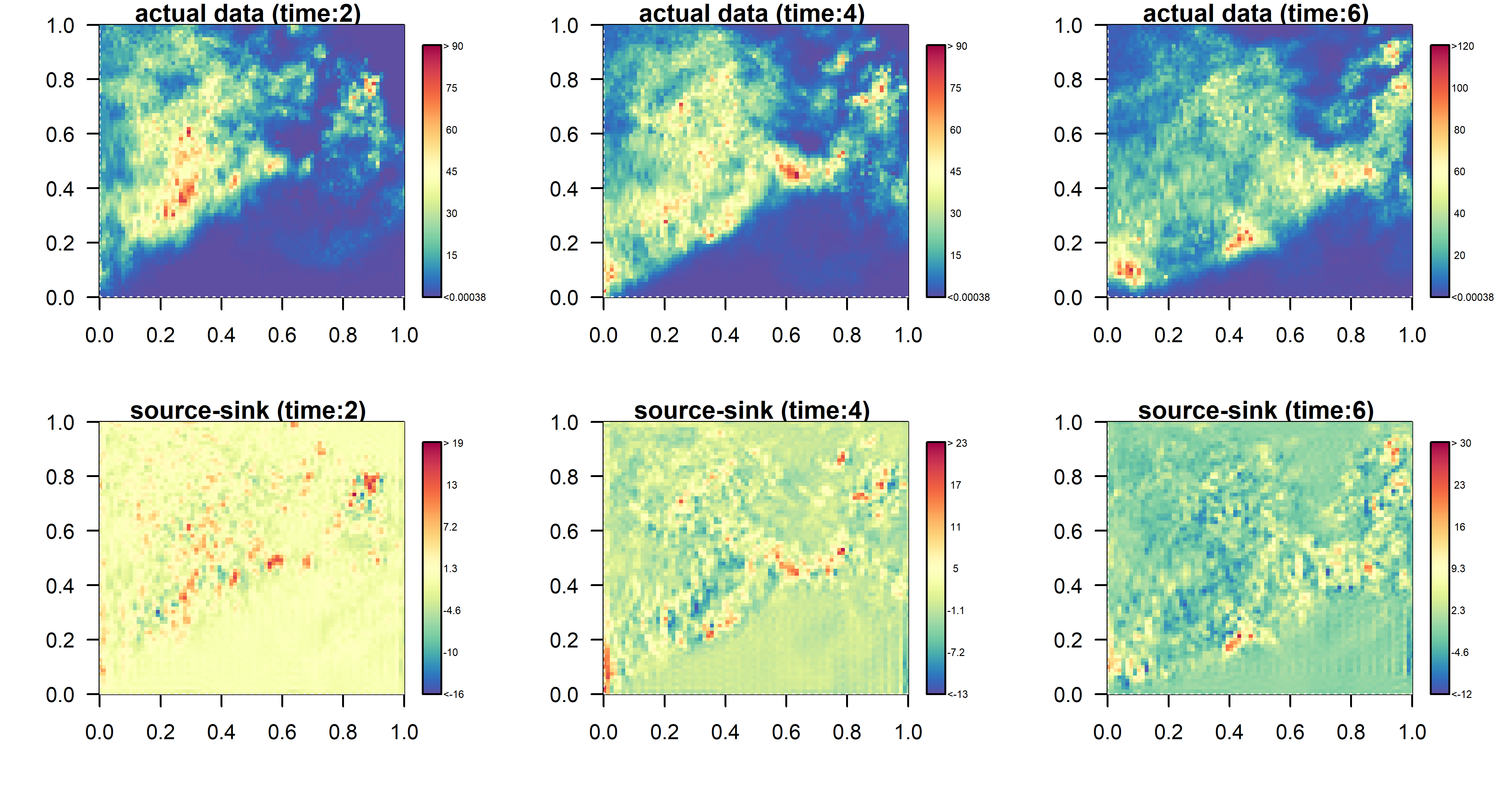}
		\centering
		\caption{Observed images (top) and estimated growth-dissipation (bottom) of the weather system in space and at selected times, $t=2,4,6$.}
		\label{fig:case_growth} 
	\end{center}
\end{figure}

For tropical thunderstorms, the growth and decay of radar reflectivity become more prominent due to the presence of many small-scale localized convective storm cells embedded in the storm systems.
The strong solar heating of land areas in tropical areas causes a phenomenon known as the convective heating, which causes the land areas to become heated more than its surroundings, and leads to significant evaporation that creates the small-scale localized convective weather cells. For such a weather system, heavy thunderstorms can develop, grow, and dissipate very suddenly in a random manner. The lifetime of convective thunderstorm cells can be as short as tens of minutes, posing a tremendous challenge to the spatio-temporal modeling of radar reflectivity data.
It has been pointed out that the errors in the linear extrapolation of radar echo fields assuming a persistent reflectivity level are mainly due to the growth and decay of reflectivity \citep{Browning1982}. 
In addition, the rapid growth-decay of weather systems also makes the NWP models ineffective in predicting the exact location and intensity of individual thunderstorms \citep{RMI2008}. The proposed spatio-temporal model, on the other hand, has the capability to capture the growth-decay of radar reflectivity in space and time. Note that, the proposed approach is motivated from the SPDE (\ref{eq:SPDE1}) with a source-sink component $Q(\bm{s},t)$, and the dynamical model (\ref{eq:dynamical_spectral}) incorporates the source-sink component as an AR(1) process. Figure \ref{fig:case_growth} shows the estimated spatial growth-dissipation of the weather system at selected times, $t=2,4,6$. These images are reconstructed by the inverse Fourier transform using the filtered $\bm{\theta}(t)$. 

\begin{figure}[h!]
	\begin{center}
		\includegraphics[width=1\textwidth]{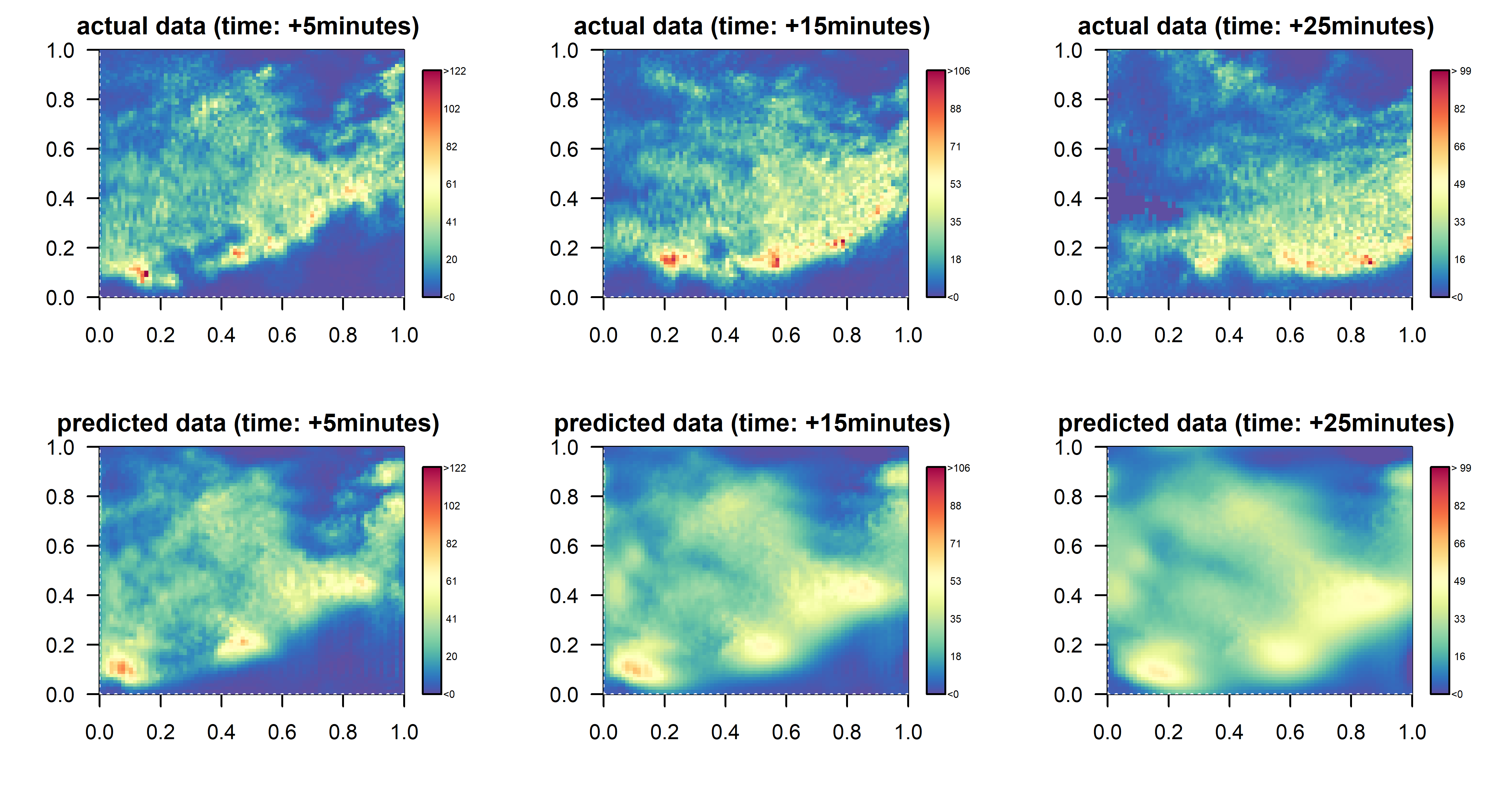}
		\centering
		\caption{Actual and predicted images for 5-, 15-, and 25-minute-ahead nowcasting.}
		\label{fig:case_pred} 
	\end{center}
\end{figure}

The dynamical model (\ref{eq:dynamical_spectral}) also enables us to perform short-term spatio-temporal extrapolation of the radar reflectivity, i.e., short-term precipitation nowcasting. Note that, in tropical areas, radar images are not used for mid-range or long-term predictions as tropical thunderstorm systems develop quickly and are of very short lifespan. For example, the official lead time for heavy storm warnings is usually between 15 and 45 minutes in tropical countries such as Singapore \citep{NEA2017}. Figure \ref{fig:case_pred} shows both the actual and 5-, 15- and 25-minute-ahead nowcasting based on our model. The spatio-temporal extrapolation of the weather system reasonably matches with the observed reflectivity fields. The nowcasting accuracy gradually deteriorates as the prediction horizon increases, which is expected given the chaotic nature of tropical weather systems.

Comparison studies are performed between the proposed method and two other methods in the literature. 
The first method is a well-known radar-based short-term precipitation prediction approach known as COTREC (Tracking Radar Echoes by Correlation), and has been widely implemented by the meteorological community \citep{Li2004}. COTREC is a typical pattern-based method based on the concept of area tracking. Tracking areas (image pixel arrays) are defined around all pixel grid points, and corresponding areas are searched in the next image by maximizing the cross-correlation between areas. Then, the velocity field can be constructed given the spatial lags between areas and the time lag between two images. Unlike the first-order-variation-based methods (e.g. Optical Flow \citep{Horn1981}), pattern-based methods assume that the shape of image brightness patterns within defined areas do not change over short time intervals.
The second method in the comparison study is a recently proposed spatio-temporal conditional autoregressive model under the Lagrangian integration scheme \citep{Liu2018}. We respectively denote these two methods by COTREC2004 and LGK2018.

\begin{table} [h!]
	\centering
	\caption{MSE for 5- to 30-minute-ahead nowcasting based on 199 storm events}
	\label{tbl:case-comparsion}
	\begin{tabular}{ p{4.2cm} | p{1.5cm} | p{1.5cm}  |p{1.5cm} |p{1.5cm} |p{1.5cm}  |p{1.5cm}   }
		\hline
		Methods & 5-min-ahead & 10-min-ahead & 15-min-ahead & 20-min-ahead & 25-min-ahead & 30-min-ahead\\ 
		\hline
		The proposed approach (spatially-varying $\mathcal{A}$) & \textbf{8.015} & \textbf{11.220} & \textbf{16.619} & \textbf{19.208 }& \textbf{26.739} & \textbf{32.936}\\ 
		\hline
		The proposed approach (spatially-invariant $\mathcal{A}$) & 8.223 & 13.721& 20.127 & 25.930 & 34.027 & 41.255\\ 
		\hline
		LGK2018 & 8.769 & 13.045 & 18.281 & 23.287 & 29.413 & 35.346\\ 
		\hline
		COTREC2004 & 8.301& 13.062 & 18.157 & 22.136 & 28.462 & 34.877 \\ 
		\hline
	\end{tabular}
	
\end{table}
Table \ref{tbl:case-comparsion} presents the comparison between the proposed methods (assuming both spatially-varying and spatially-invariant convection-diffusion operators), COTREC2004 and LGK2018. The Mean-Squared-Error (MSE) is reported for 5-minute- to 30-minute-ahead nowcasting 
of the (low-pass) radar images based on \emph{199} tropical storm events recorded in 2010 and 2011.
It is seen that the proposed method assuming spatially-varying convection-diffusion outperforms the same approach assuming spatially-invariant convection-diffusion, COTREC2004, and LGK2018. This advantage is primarily due to the improvement in estimating the wind field by the proposed method. Both COTREC2004 and LGK2018 estimate the wind field using the pattern-based methods that require the shape of radar images do not change over short time intervals within the defined tracking areas. This assumption is often violated by highly dynamic tropical thunderstorms. In addition, the assumption of spatially-invariant convection-diffusion is obviously inappropriate, causing a rapid increase of MSE as the nowcasting horizon grows. Such observations demonstrate the significance of incorporating spatially-varying convection-diffusion in this work.


\subsubsection{Discussions on the Periodicity Assumption}
When the proposed method is implemented, it is important to note that the choice of the Fourier basis functions is only optimal when the process $\xi(t,\bm{s})$ is periodic. Hence, for the three reasons discussed in Section \ref{sec:data_case}, it is always necessary to embed the spatial area of interest into a larger computational spatial domain. In this section, we provide the readers with additional insights on the effect of the periodicity assumption, if the spatial domain of interest is not embedded into a larger computational domain. 

At a chosen point of time $t_0$, Figure 11 shows the actual (top row) and forecast (bottom row) weather radar images after 5, 20 and 35 minutes. The time $t_0$ is so chosen that the weather system is moving out of the boundary of the spatial domain from the southeast corner (as seen in the first row). However, due to the periodicity assumption, the predicted weather system moves back into the spatial domain as indicated by the red arrows. This numerical example strongly justifies the importance of embedding the spatial domain of interest into a bigger computational domain when the proposed approach is adopted. In fact, the extrapolation nature of radar-based precipitation forecast determines that it is impossible to predict the radar reflectivity at the influx boundaries of the spatial domain. 
\begin{figure}[h!]  
	\label{fig:edge}
	\begin{center}  
		\includegraphics[width=1\textwidth]{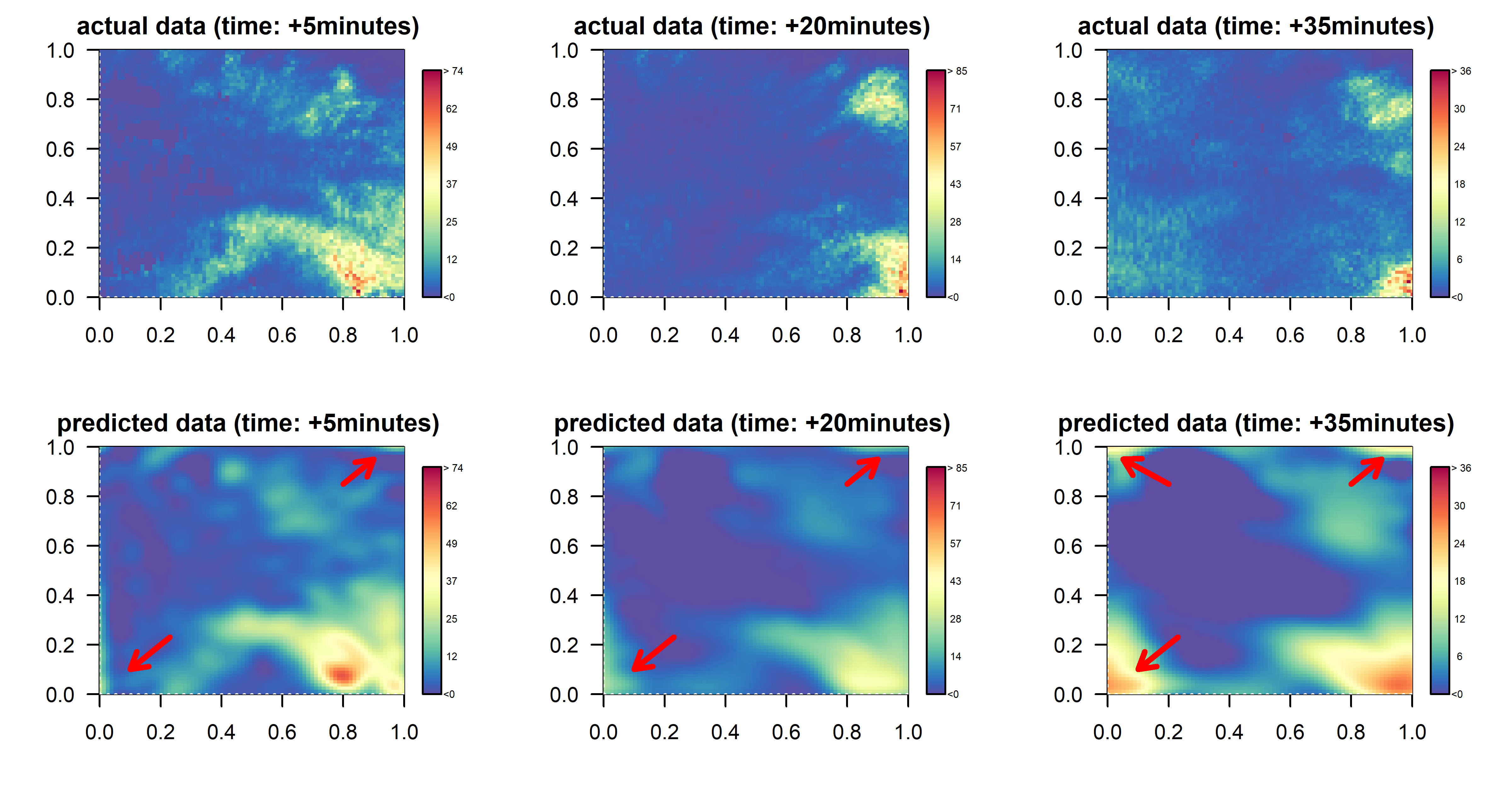} 
		\centering
		\caption{Illustration of the effect of the periodicity assumption when the spatial domain of interest is not embedded into a bigger computational domain. The top and bottom rows, respectively, show the actual and predicted radar images after 5, 20 and 35 minutes. The red arrows indicate that the predicted weather system moves back into the spatial domain precisely due to the periodicity assumption, while the actual weather system leaves the spatial domain.}
	\end{center}
\end{figure}

\section{Conclusions}
This paper proposed a statistical modeling approach for spatio-temporal data arising from a class of physical convection-diffusion processes. 
Motivated by existing results in the literature, the proposed approach is rooted in the spectrum decomposition of spatio-temporal processes. However, unlike existing results, this paper obtained the statistical model by considering spatially-varying convection-diffusion and nonzero-mean source-sink. Due to the spatially-varying convection-diffusion, the
temporal evolution of individual spectrum coefficients is coupled, which extends the existing results to a more complicated but realistic scenario. This phenomenon is known as the non-linear energy transfer across multiple scales due to spatially-varying convection-diffusion. As a result, the proposed spatio-temporal model has a non-stationary covariance structure which has been established by our theoretical investigation. A dynamical model, in the spectral domain, has been constructed and the conditional structure of the dynamical model leads to tremendous advantages in statistical inference and computation. The proposed approach has been illustrated by a numerical example as well as a case study based on real datasets. The advantages of the proposed method have been demonstrated through a comparison study involving 199 storm events. Note that, since the convection-diffusion processes can be widely found in interdisciplinary domains including environmental sciences, image analysis, geology, material science, reliability engineering, etc., the proposed method has the potential to impact a spectrum of scientific and engineering applications. The R code is available at GitHub: \url{https://github.com/dnncode/Spatio-Temporal-Model-for-SPDE}.

\section{Appendices}
\appendix
\section{Proof of Proposition 1}
Since $\bm{\vec{v}}_{\bm{s}}^T\triangledown f_j(\bm{s})=\imath \bm{\vec{v}}_{\bm{s}}^T \bm{k}_j f_j(\bm{s})$ and $\triangledown \cdot [\bm{D}_{\bm{s}} \triangledown f_j(\bm{s})]=(-\bm{k}_j^T \bm{D}_{\bm{s}}\bm{k}_j + \imath  [\triangledown \cdot \bm{D}_{\bm{s}}]^T \bm{k}_j )f_j(\bm{s})$, substituting the spectral decomposition (\ref{eq:decomposition}) into the SPDE (\ref{eq:SPDE1}) yields:
\begin{equation} \label{eq:proof_1_1}
\begin{split}
\sum_{j}^{K}f_j(\bm{s})\dot{\alpha}_j(t) = &  -\sum_{j}^{K} \imath  \bm{\vec{v}}_{\bm{s}}^T \bm{k}_j f_j(\bm{s}) \alpha_j(t)  + \sum_{j}^{K}(-\bm{k}_j^T \bm{D}_{\bm{s}}\bm{k}_j + \imath [\triangledown \cdot \bm{D}_{\bm{s}}]^T \bm{k}_j ) f_j(\bm{s}) \alpha_j(t) \\ & - \zeta_{\bm{s}} \sum_{j}^{K}f_j(\bm{s})\alpha_j(t)  +\sum_{j}^{K}f_j(\bm{s})\beta_j(t) + \sum_{j}^{K}f_j(\bm{s})\tilde{\varepsilon}_j(t).
\end{split}
\end{equation} 

Using the Galerkin method, we multiply both the left hand side (LHS) and right hand side (RHS) of (\ref{eq:proof_1_1}) by the complex conjugate $f_l^*(\bm{s})$ ($l=1,2,...,K$), and integrate the product over the spatial domain. Because the Fourier basis functions are orthogonal, i.e., $\int f_j(\bm{s})f_l^*(\bm{s})d\bm{s}=0$ if $j \neq l$, the LHS of (\ref{eq:proof_1_1}) becomes
\begin{equation} 
\int_{\mathbb{S}}\sum_{j}^{K}f_j(\bm{s})f_l^*(\bm{s})\dot{\alpha}_j(t)d\bm{s} = \dot{\alpha}_l(t) \int_{\mathbb{S}} f_j(\bm{s})f_j^*(\bm{s})d\bm{s} \equiv \dot{\alpha}_l(t) C_j
\end{equation}
for $l=1,2,...,K$.

Similarly, for any $l=1,2,...,K$, we multiply the complex conjugate $f_l^*(\bm{s})$ to the RHS of (\ref{eq:proof_1_1}) and integrate the product over the spatial domain. Then, the five terms on the RHS are respectively given by
\begin{equation} 
\int_{\mathbb{S}} -\sum_{j}^{K} \imath  \bm{\vec{v}}_{\bm{s}}^T \bm{k}_j f_j(\bm{s}) f_l^*(\bm{s}) \alpha_j(t) d\bm{s} = \sum_{j}^{K}\left(-\imath  \int \bm{\vec{v}}_{\bm{s}}^T \bm{k}_j f_j(\bm{s})f_l^*(\bm{s}) d\bm{s} \right)\alpha_j(t) \equiv \sum_{j}^{K}g_{l,j}^{(1)}\alpha_j(t)
\end{equation}
\begin{equation} 
\begin{split}
\int_{\mathbb{S}} \sum_{j}^{K}(-\bm{k}_j^T \bm{D}_{\bm{s}}\bm{k}_j + \imath [\triangledown \cdot \bm{D}_{\bm{s}}]^T \bm{k}_j ) & f_j(\bm{s})f_l^*(\bm{s}) \alpha_j(t) d\bm{s} \\ & = 
\sum_{j}^{K}\left(\int (-\bm{k}_j^T \bm{D}_{\bm{s}}\bm{k}_j + \imath [\triangledown \cdot \bm{D}_{\bm{s}}]^T \bm{k}_j ) f_j(\bm{s}) f_l^*(\bm{s}) d\bm{s}  \right)\alpha_j(t) \\ & \equiv \sum_{j}^{K}g_{l,j}^{(2)}\alpha_j(t)
\end{split}
\end{equation}
\begin{equation} 
\int_{\mathbb{S}} \sum_{j}^{K} \zeta_{\bm{s}}  f_j(\bm{s})f_l^*(\bm{s}) \alpha_j(t) d\bm{s} = \sum_{j}^{K} \left( \int \zeta_{\bm{s}}  f_j(\bm{s})f_l^*(\bm{s}) d\bm{s} \right) \alpha_j(t) \equiv \sum_{j}^{K}g_{l,j}^{(3)}\alpha_j(t)
\end{equation}
\begin{equation} 
\int_{\mathbb{S}} \sum_{j}^{K}f_j(\bm{s})f_l^*(\bm{s})\beta_j(t) d\bm{s} = C_j \beta_l(t) 
\end{equation}
\begin{equation} 
\int_{\mathbb{S}} \sum_{j}^{K}f_j(\bm{s})f_l^*(\bm{s})\tilde{\varepsilon}_j(t) d\bm{s} = C_j \tilde{\varepsilon}_l(t) 
\end{equation}

Let $g_{i,j} = C_j^{-1} (g_{i,j}^{(1)}+g_{i,j}^{(2)}+g_{i,j}^{(3)})$, we obtain
\begin{equation} 
\dot{\alpha}_l(t) =  \sum_{j}^{K} g_{l,j} \alpha_j(t) + \beta_l(t) + \tilde{\varepsilon}_l(t)
\end{equation}
which implies $	\dot{\bm{\alpha}}(t) =  \bm{G}\bm{\alpha}(t) + \bm{\beta}(t) + \bm{\tilde{\varepsilon}}(t)$, as was to be proved $\blacksquare$
\vspace{0.2in}

\section{Proof of Proposition 2}
Solving the ODE (\ref{eq:gamma_transition_2}) for discrete time points yields
\begin{equation}  \label{eq:alpha_solved}
\bm{\alpha}(t+\Delta)  = \exp(\bm{G}\Delta)\bm{\alpha}(t) + \bm{q}(\Delta)
\end{equation}
where
\begin{equation} \label{eq:alpha_solved_2}
\bm{q}(\Delta) \sim N \left(  \int_{t}^{t+\Delta}\bm{\beta}(t)dt, \int_{0}^{\Delta}\exp(\bm{G}(\Delta-\tau))\bm{H}\exp^*(\bm{G}^T(\Delta-\tau))d\tau\right).
\end{equation}

Hence, the covariance matrix of $\bm{\alpha}(t)$ is
\begin{equation} \label{eq:p2_proof2}
\mathrm{var}(\bm{\alpha}(t)) = \exp(\bm{G}t)\bm{H}_0\exp^*(\bm{G}^Tt) + \int_{0}^{t}\exp(\bm{G}(t-\tau))\bm{H}\exp^*(\bm{G}^T(t-\tau))d\tau.
\end{equation}

It is also noted that
\begin{equation}  \label{eq:p2_proof1}
\begin{split}
\mathrm{cov}(\bm{\alpha}(t+\Delta), \bm{\alpha}(t)) & = \mathrm{cov}(\exp(\bm{G}\Delta)\bm{\alpha}(t) + \bm{q}(\Delta), \bm{\alpha}(t)) \\ & = \exp(\bm{G}\Delta) \mathrm{var}(\bm{\alpha}(t)).
\end{split}
\end{equation}

Substituting (\ref{eq:p2_proof2}) into (\ref{eq:p2_proof1}) immediately implies (\ref{eq:alpha_cov}).

\section{Proof of proposition 3}
Note that, $\bm{\vec{v}}_{\bm{s}}^T\triangledown f_{\bm{k}}^{(R)}(\bm{s})=-\bm{\vec{v}}_{\bm{s}}^T \bm{\tilde{k}} f_{\bm{k}}^{(I)}(\bm{s})$, 
$\bm{\vec{v}}_{\bm{s}}^T\triangledown f_{\bm{k}}^{(I)}(\bm{s})= \bm{\vec{v}}_{\bm{s}}^T \bm{\tilde{k}} f_{\bm{k}}^{(R)}(\bm{s})$,
$\triangledown \cdot [\bm{D}_{\bm{s}} \triangledown f_{\bm{k}}^{(R)}(\bm{s})]=-\bm{\tilde{k}}^T \bm{D}_{\bm{s}}\bm{\tilde{k}}f_{\bm{k}}^{(R)}(\bm{s}) - [\triangledown \cdot \bm{D}_{\bm{s}}]^T \bm{\tilde{k}} f_{\bm{k}}^{(I)}(\bm{s})$, 
and $\triangledown \cdot [\bm{D}_{\bm{s}} \triangledown f_{\bm{k}}^{(I)}(\bm{s})]=-\bm{\tilde{k}}^T \bm{D}_{\bm{s}}\bm{\tilde{k}}f_{\bm{k}}^{(I)}(\bm{s}) + [\triangledown \cdot \bm{D}_{\bm{s}}]^T \bm{\tilde{k}} f_{\bm{k}}^{(R)}(\bm{s})$.
Then, based on the Galerkin projection, we respectively multiply the LHS of the SPDE (\ref{eq:SPDE1}) by $f_{\bm{k}'}^{(R)}(\bm{s})$ and $f_{\bm{k}'}^{(I)}(\bm{s})$ for $\bm{k}' \in \Omega_1 \cup \Omega_2$, and integrate the product over the spatial domain. Then, the LHS of (\ref{eq:SPDE1}) becomes
\begin{equation} 
\begin{split}
& \int_{\mathbb{S}}\frac{\partial}{\partial t} \xi(t,\bm{s}) f_{\bm{k}'}^{(R)}(\bm{s}) d\bm{s} \\
&= \int_{\mathbb{S}} \left(  \sum_{\bm{k}\in\Omega_1}\dot{\alpha}_{\bm{k}}^{(R)}(t)f_{\bm{k}}^{(R)}(\bm{s})
+ 2\sum_{\bm{k}\in\Omega_2}(\dot{\alpha}_{\bm{k}}^{(R)}(t)f_{\bm{k}}^{(R)}(\bm{s})+\dot{\alpha}_{\bm{k}}^{(I)}(t)f_{\bm{k}}^{(I)}(\bm{s})) \right) f_{\bm{k}'}^{(R)}(\bm{s}) d\bm{s} \\
& =  \begin{cases}
C_{\bm{k}}\dot{\alpha}_{\bm{k}'}^{(R)}(t) \quad\quad \text{for $\bm{k}' \in \Omega_1$} \\
2C_{\bm{k}}\dot{\alpha}_{\bm{k}'}^{(R)}(t) \quad\quad  \text{for $\bm{k}' \in \Omega_2$} 
\end{cases}
\end{split}
\end{equation}
and
\begin{equation} 
\begin{split}
& \int_{\mathbb{S}}\frac{\partial}{\partial t} \xi(t,\bm{s}) f_{\bm{k}'}^{(I)}(\bm{s}) d\bm{s} \\
&= \int_{\mathbb{S}} \left(  \sum_{\bm{k}\in\Omega_1}\dot{\alpha}_{\bm{k}}^{(R)}(t)f_{\bm{k}}^{(R)}(\bm{s})
+ 2\sum_{\bm{k}\in\Omega_2}(\dot{\alpha}_{\bm{k}}^{(R)}(t)f_{\bm{k}}^{(R)}(\bm{s})+\dot{\alpha}_{\bm{k}}^{(I)}(t)f_{\bm{k}}^{(I)}(\bm{s})) \right) f_{\bm{k}'}^{(I)}(\bm{s}) d\bm{s} \\
& =  \begin{cases}
0 \quad\quad \text{for $\bm{k}' \in \Omega_1$} \\
2C_{\bm{k}}\dot{\alpha}_{\bm{k}'}^{(I)}(t) \quad\quad  \text{for $\bm{k}' \in \Omega_2$} 
\end{cases}
\end{split}
\end{equation}

Similarly, we respectively multiply the RHS of the SPDE (\ref{eq:SPDE1}) by $f_{\bm{k}'}^{(R)}(\bm{s})$ and $f_{\bm{k}'}^{(I)}(\bm{s})$ for $\bm{k}' \in \Omega_1 \cup \Omega_2$, and integrate the product over the spatial domain. Then, the first term $-\bm{\vec{v}}_{\bm{s}}^T \triangledown \xi(t,\bm{s})$ on the RHS of (\ref{eq:SPDE1}) becomes:
\begin{equation} 
-\bm{\vec{v}}_{\bm{s}}^T \triangledown \xi(t,\bm{s}) = \sum_{\bm{k}\in\Omega_1}\bm{\vec{v}}_{\bm{s}}^T\tilde{\bm{k}}f_{\bm{k}}^{(I)}(\bm{s})\alpha_{\bm{k}}^{(R)}(t) + 2\sum_{\bm{k}\in\Omega_2}\left\{ \bm{\vec{v}}_{\bm{s}}^T\tilde{\bm{k}}f_{\bm{k}}^{(I)}(\bm{s})\alpha_{\bm{k}}^{(R)}(t) - \bm{\vec{v}}_{\bm{s}}^T\tilde{\bm{k}}f_{\bm{k}}^{(R)}(\bm{s})\alpha_{\bm{k}}^{(I)}(t)\right\}.
\end{equation}

Hence, for any $\bm{k}' \in \Omega_1 \cup \Omega_2$, we have
\begin{equation} 
-\int_{\mathbb{S}}\bm{\vec{v}}_{\bm{s}}^T \triangledown \xi(t,\bm{s}) f_{\bm{k}'}^{(R)}(\bm{s}) d\bm{s} 
= \sum_{\bm{k}\in\Omega_1} \alpha_{\bm{k}}^{(R)}(t)  \Psi_1(\bm{k},\bm{k}') 
+ 2\sum_{\bm{k}\in\Omega_2}\left\{\alpha_{\bm{k}}^{(R)}(t)  \Psi_1(\bm{k},\bm{k}')
+ \alpha_{\bm{k}}^{(I)}(t)\Psi_2(\bm{k},\bm{k}')\right\}
\end{equation}
\begin{equation} 
-\int_{\mathbb{S}}\bm{\vec{v}}_{\bm{s}}^T \triangledown \xi(t,\bm{s}) f_{\bm{k}'}^{(I)}(\bm{s}) d\bm{s} 
= \sum_{\bm{k}\in\Omega_1} \alpha_{\bm{k}}^{(R)}(t)  \Psi_3(\bm{k},\bm{k}') 
+ 2\sum_{\bm{k}\in\Omega_2}\left\{\alpha_{\bm{k}}^{(R)}(t)  \Psi_3(\bm{k},\bm{k}')
+ \alpha_{\bm{k}}^{(I)}(t)\Psi_4(\bm{k},\bm{k}')\right\}
\end{equation}

The second term $\triangledown \cdot [\bm{D}_{\bm{s}} \triangledown \xi(t,\bm{s})]$ on the RHS of (\ref{eq:SPDE1}) becomes:
\begin{equation} 
\begin{split}
\triangledown \cdot [\bm{D}_{\bm{s}} \triangledown \xi(t,\bm{s})] = & 
\sum_{\bm{k}\in\Omega_1}\alpha_{\bm{k}}^{(R)}(t)(-\tilde{\bm{k}}^T \bm{D}_{\bm{s}} \tilde{\bm{k}}  f_{\bm{k}}^{(R)}-  [\triangledown \cdot \bm{D}_{\bm{s}}]^T \tilde{\bm{k}}  f_{\bm{k}}^{(I)} ) \\
& + 2\sum_{\bm{k}\in\Omega_2}\alpha_{\bm{k}}^{(R)}(t)(-\tilde{\bm{k}}^T \bm{D}_{\bm{s}} \tilde{\bm{k}}  f_{\bm{k}}^{(R)}-  [\triangledown \cdot \bm{D}_{\bm{s}}]^T \tilde{\bm{k}}  f_{\bm{k}}^{(I)} ) \\
& + 2\sum_{\bm{k}\in\Omega_2}\alpha_{\bm{k}}^{(I)}(t)(-\tilde{\bm{k}}^T \bm{D}_{\bm{s}} \tilde{\bm{k}}  f_{\bm{k}}^{(I)}-  [\triangledown \cdot \bm{D}_{\bm{s}}]^T \tilde{\bm{k}}  f_{\bm{k}}^{(R)} ). 
\end{split}
\end{equation}

Hence, for any $\bm{k}' \in \Omega_1 \cup \Omega_2$, we have
\begin{equation} 
\int_{\mathbb{S}} \triangledown \cdot [\bm{D}_{\bm{s}} \triangledown \xi(t,\bm{s})] f_{\bm{k}'}^{(R)}(\bm{s}) d\bm{s} 
= \sum_{\bm{k}\in\Omega_1} \alpha_{\bm{k}}^{(R)}(t)  \Psi_5(\bm{k},\bm{k}') 
+ 2\sum_{\bm{k}\in\Omega_2}\left\{\alpha_{\bm{k}}^{(R)}(t)  \Psi_5(\bm{k},\bm{k}')
+ \alpha_{\bm{k}}^{(I)}(t)\Psi_6(\bm{k},\bm{k}')\right\}
\end{equation}
\begin{equation} 
\int_{\mathbb{S}} \triangledown \cdot [\bm{D}_{\bm{s}} \triangledown \xi(t,\bm{s})] f_{\bm{k}'}^{(I)}(\bm{s}) d\bm{s} 
= \sum_{\bm{k}\in\Omega_1} \alpha_{\bm{k}}^{(R)}(t)  \Psi_7(\bm{k},\bm{k}') 
+ 2\sum_{\bm{k}\in\Omega_2}\left\{\alpha_{\bm{k}}^{(R)}(t)  \Psi_7(\bm{k},\bm{k}')
+ \alpha_{\bm{k}}^{(I)}(t)\Psi_8(\bm{k},\bm{k}')\right\}.
\end{equation}

The third term $-\zeta_{\bm{s}} \xi(t,\bm{s})$ on the RHS of (\ref{eq:SPDE1}) becomes:
\begin{equation} 
-\zeta_{\bm{s}} \xi(t,\bm{s}) = - \zeta_{\bm{s}} \left( \sum_{\bm{k}\in\Omega_1}\alpha_{\bm{k}}^{(R)}(t)f_{\bm{k}}^{(R)}(\bm{s})
+ 2\sum_{\bm{k}\in\Omega_2}(\alpha_{\bm{k}}^{(R)}(t)f_{\bm{k}}^{(R)}(\bm{s})+\alpha_{\bm{k}}^{(I)}(t)f_{\bm{k}}^{(I)}(\bm{s})) \right).
\end{equation}
Hence, for any $\bm{k}' \in \Omega_1 \cup \Omega_2$, we have
\begin{equation} 
\int_{\mathbb{S}}-\zeta_{\bm{s}} \xi(t,\bm{s}) f_{\bm{k}'}^{(R)}(\bm{s})d\bm{s} 
= \sum_{\bm{k}\in\Omega_1} \alpha_{\bm{k}}^{(R)}(t)  \Psi_9(\bm{k},\bm{k}') 
+ 2\sum_{\bm{k}\in\Omega_2}\left\{\alpha_{\bm{k}}^{(R)}(t)  \Psi_9(\bm{k},\bm{k}')
+ \alpha_{\bm{k}}^{(I)}(t)\Psi_{10}(\bm{k},\bm{k}')\right\}.
\end{equation}
\begin{equation} 
\int_{\mathbb{S}}-\zeta_{\bm{s}} \xi(t,\bm{s}) f_{\bm{k}'}^{(I)}(\bm{s})d\bm{s} 
= \sum_{\bm{k}\in\Omega_1} \alpha_{\bm{k}}^{(R)}(t)  \Psi_{11}(\bm{k},\bm{k}') 
+ 2\sum_{\bm{k}\in\Omega_2}\left\{\alpha_{\bm{k}}^{(R)}(t)  \Psi_{11}(\bm{k},\bm{k}')
+ \alpha_{\bm{k}}^{(I)}(t)\Psi_{12}(\bm{k},\bm{k}')\right\}.
\end{equation}
as was to be proved $\blacksquare$

\section{Proof of Proposition 4}
For a spatially-varying but temporally-invariant convection-diffusion operator $\mathcal{A}$, the convection-diffusion operation $\mathcal{A}\tilde{\xi}(t,\bm{s})$ on the approximated process $\tilde{\xi}(t,\bm{s}) = \bm{f}^T(\bm{s})\bm{\alpha}(t)$ leads to 
\begin{eqnarray} \label{eq:p4_proof1}
\bm{\alpha}(t+\Delta) =  \exp(\bm{G}\Delta)\bm{\alpha}(t-1) \equiv \mathcal{\bm{G}}\bm{\alpha}(t-1)
\end{eqnarray}
for discrete time steps with lag $\Delta$. Note that, $\bm{\alpha}(t)$ is a vector of Fourier coefficients at frequencies $\bm{k}_1, \bm{k}_2, ..., \bm{k}_N$. Here, we explicitly write $\bm{\alpha}(t)$ as $\bm{\alpha}(t) = (\alpha_{\bm{k}_1}(t), \alpha_{\bm{k}_2}(t), ..., \alpha_{\bm{k}_N}(t))^T$.  

Applying discrete inverse Fourier transform to both sides of (\ref{eq:p4_proof1}), the left side becomes
\begin{eqnarray} 
\mathcal{F}^{-1}(\bm{\alpha}(t+\Delta)) = \sum_{j=1}^{N} \alpha_{\bm{k}_j}(t+\Delta) e^{\imath \bm{k}_j^T \bm{s}}=\tilde{\xi}(t+\Delta,\bm{s}), 
\end{eqnarray}
while the right side of (\ref{eq:p4_proof1}) yields:
\begin{eqnarray}  \label{eq:RHS}
\mathcal{F}^{-1}(\mathcal{G}\bm{\alpha}(t) ) = \sum_{j=1}^{N}\sum_{j'=1}^{N}\mathcal{G}_{j,j'}\alpha_{\bm{k}_{j'}}(t)e^{\imath \bm{k}_j^T \bm{s}}
\end{eqnarray}
where $\mathcal{G}_{i,j}$ is the $(j,j')$th entry of the matrix $\mathcal{G}$. 

Note that, $\bm{\alpha}_{\bm{k}_{j'}}(t) = N^{-1}\sum_{j=1}^{N}\tilde{\xi}(t,\bm{x}_i)e^{-\imath \bm{k}_{j'}^T \bm{x}_i}$ from the discrete Fourier transform. Hence, (\ref{eq:RHS}) can be written as
\begin{eqnarray}  \label{eq:RHS2}
\begin{split}
\sum_{j=1}^{N}\sum_{j'=1}^{N}\mathcal{G}_{j,j'}\alpha_{\bm{k}_{j'}}(t)e^{\imath \bm{k}_j^T \bm{s}} & =  \frac{1}{N} \sum_{i=1}^{N} \left[\sum_{j=1}^{N} \sum_{j'=1}^{N}  \mathcal{G}_{j,j}e^{\imath \bm{k}_j\bm{s} - \imath\bm{k}_{j'}\bm{x}_i}\right] \tilde{\xi}(t,\bm{x}_i) \\
& = \frac{1}{N} \sum_{i=1}^{N} \omega_{\bm{s}}(\bm{x}_i) \tilde{\xi}(t,\bm{x}_i)
\end{split}
\end{eqnarray}
and
\begin{eqnarray} \label{eq:p4_proof2}
\tilde{\xi}(t+\Delta,\bm{s}) = \frac{1}{N} \sum_{i=1}^{N} \omega_{\bm{s}}(\bm{x}_i) \tilde{\xi}(t,\bm{x}_i)
\end{eqnarray}
as was to be shown $\blacksquare$

\section{Discussions on the Computation Aspects}
For spatially-varying convection, the computational bottle neck of the proposed model is the computation of a $N_1N_2 \times N_1N_2$ dense and non-symmetric matrix $\bm{G}$ in (4). The evaluation of each component of $\bm{G}$ requires the numerical integration of complex functions over the spatial domain. Note that, when the convection-diffusion does not vary in space and time, $\bm{G}$ reduces to a diagonal matrix and the evaluation of each component of $\bm{G}$ does not require numerical integration. 

For example, our case study considers radar images with $200\times200$ pixels. Hence, $\bm{G}$ is a $40,000 \times 40,000$ matrix, and the evaluation of each element of $\bm{G}$, i.e., $g_{i,j}$ in (\ref{eq:gamma_transition_2}), requires a numerical integration of complex functions over the spatial domain. A total number of $1.6 \times 10^9$ numerical integrations are needed for evaluating the dense and non-symmetric matrix $\bm{G}$. Since $\bm{G}$ contains unknown parameters, the statistical inference requires $\bm{G}$ to be computed repeatedly, which is computationally expensive. 

\begin{figure}[h!]  
	\begin{center}
		\includegraphics[width=0.9\textwidth]{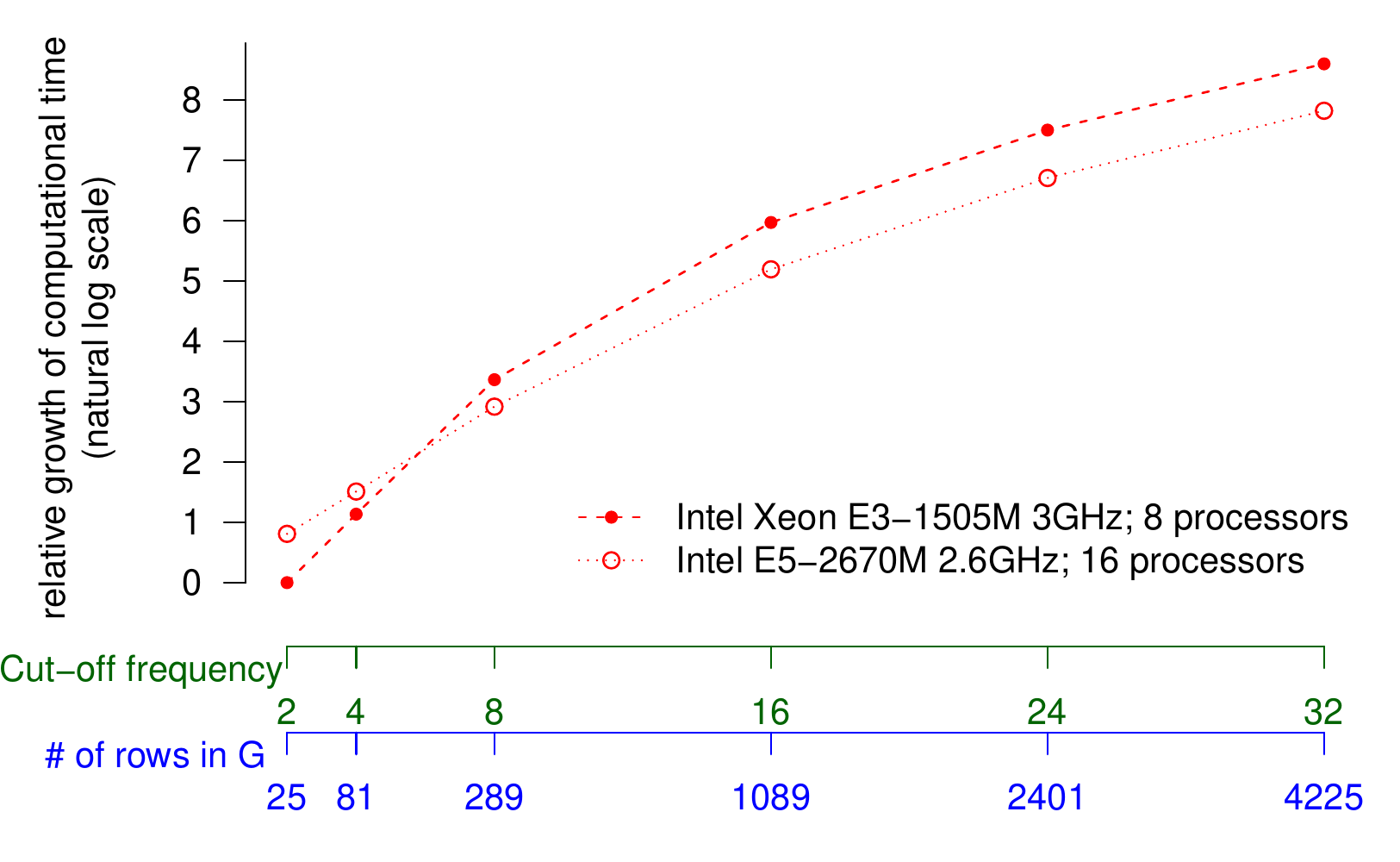}
		\centering
		\caption{Computational scalability for evaluating the matrix $\bm{G}$}
		\label{fig:scale}
	\end{center}
\end{figure}

Figure \ref{fig:scale} provides the readers with some insights on the scalability of the computation of $\bm{G}$ as the dimension of $\bm{G}$ grows. 
In this figure, the first horizontal axis shows the cut-off frequency for low-pass filtering. For example, if the cut-off frequency equals 4, then, only those frequency components with their (absolute) angular wavenumber less than or equal 4 radians per unit distance are retained. The second horizontal axis shows the dimension of $\bm{G}$ corresponding to each chosen cut-off frequency. The vertical axis shows the relative growth of computational time. The computational time for a cut-off frequency of 2 radians per unit distance is used as the baseline for calculating the relatively growth in computational time. 

Our code is written in R, and the evaluation of the components in $\bm{G}$ are performed in parallel using the \texttt{foreach} packages (i.e., parallel for loop). The code is implemented on a Dell workstation with 8 logical processors (Intel Xeon E3-1505M v6 3GHz), as well as the High Performance Computer (HPC) facilities with 16 nodes (Intel E5-2670 2.6GHz).

Because the statistical inference of the model parameters requires $\bm{G}$ to be repeatedly computed, we adopt a more practical two-step procedure in the case study in Section 5: 1) Step 1: estimate the unknown parameters (especially the wind field) using the low-pass filtered images by keeping only a relatively small number of low-frequency coefficients; 2) Step 2: perform Kalman Filter with known parameters. This two-step procedure is justified as follows: the most important parameter to be estimated is $\bm{\gamma}$ that determines the velocity field; see (\ref{eq:MeanFunction}). The (macro-scale) velocity field can be well estimated from the dynamics/transition of the low-frequency terms which dominate the large-scale dynamics of the spatio-temporal process. Hence, in Step 1, only the low-frequencies coefficients are used (which maintains a low dimension of $\bm{G}$) to estimate the unknown parameters. In our code, the evaluation of $\bm{G}$ is approximately half a second for a cut-off frequency of 4 radians per unit distance. After the unknown parameters have been estimated, the Kalman Filter with known parameters is relatively fast.

\bibliographystyle{asa}
\bibliography{references}


\end{document}